\documentclass[a4paper,fleqn,usenatbib]{mnras}

\usepackage[T1]{fontenc}
\usepackage{ae,aecompl}
\usepackage{textcomp}

\usepackage{graphicx}	
\usepackage{amsmath}	
\usepackage{amssymb}	
\usepackage{multicol}        
\usepackage{bm}		
\usepackage{pdflscape}	
\usepackage{color}          

\newcommand{\gaia}{{\it Gaia}}
\newcommand{\mz}{$M_{V}-{\rm [Fe/H]}$}
\newcommand{\gz}{$M_{G}-{\rm [Fe/H]}$}

\newcommand{\plz}{$PLZ$}
\title[A fresh look at RRLs in Draco with {\it Gaia}]{A fresh look at the RR Lyrae population in the Draco dwarf spheroidal galaxy with {\it \textbf{Gaia}}}

\author[T. Muraveva et al.]{Tatiana Muraveva$^{1}$\thanks{tatiana.muraveva@inaf.it}, Gisella Clementini$^{1}$, Alessia Garofalo$^{1}$, Felice Cusano$^{1}$
\\
$^{1}$ INAF-Osservatorio di Astrofisica e Scienza dello Spazio di Bologna, Via Piero Gobetti, 93/3, Bologna 40129, Italy\\
}

\date{Accepted . Received ; in original form }

\pubyear{2019}

\begin{document}
\label{firstpage}
\pagerange{\pageref{firstpage}--\pageref{lastpage}}
\maketitle

\begin{abstract}

We present a catalogue of 285 RR Lyrae stars (RRLs) in the Draco dwarf spheroidal galaxy (dSph), obtained by combining data from a number of different surveys including the second data release (DR2) of the European Space Agency (ESA) cornerstone mission {\gaia}. We have determined individual distances to the RRLs in our sample using for the first time a {\gaia} $G$-band luminosity-metallicity relation  ({\gz}) and study the structure of the Draco dSph as traced by its RRL population. We find that the RRLs located in the  western/south-western region of Draco appear to be closer to us, which may be a clue of interaction between Draco and the Milky Way (MW). The average  distance modulus of Draco measured with the RRLs is $\mu=19.53\pm0.07$~mag, corresponding to a distance of  $80.5\pm2.6$~kpc, in good agreement with previous determinations in the literature. Based on the pulsation properties of the RRLs we confirm the Oosterhoff-intermediate nature of Draco. We present an additional sample of  41 candidate RRLs in Draco, which  we selected from the {\gaia} DR2 catalogue based on the uncertainty of their $G$-band  magnitudes. Additional epoch data that will become available in the {\gaia} third data release (DR3) will help to confirm whether these candidates are bona-fide Draco RRLs.

\end{abstract}



\section{Introduction}\label{sec:intro}

RR Lyrae stars (RRLs) play an important  role in different branches of  astronomy.
They are radially pulsating variables that populate  the instability strip region of the horizontal branch (HB) in the colour-magnitude diagram (CMD), hence they can give a clue of the HB morphology and help characterising the core helium burning evolutionary stage of low mass stars (< 1 M$_{\sun}$). 
RRLs
 play an important role in the study of the resolved stellar population in galaxies as they are valuable tracers of the old stellar population (Age > 10 Gyr) abundant in globular clusters and galactic halos. 
Specific properties of the RRLs belonging to Local Group galaxies,  such as the Oosterhoff dichotomy \citep{Oo1939}, allow us to constrain to what extent these systems could have contributed to the formation of larger galaxies (e.g. \citealt{Clementini2009}) and, therefore, test existing cosmological models. Finally, RRLs are important distance indicators, since their luminosity/absolute magnitude (hence distance) can be inferred from the observed de-reddened apparent magnitude by means of the absolute magnitude - metallicity relation ({\mz}) in the visual band (e.g.  \citealt{Clementini2003}; \citealt{Cacciari2003}; \citealt{Bono2003}; \citealt{Clementini2017}; \citealt{Muraveva2018a}) and period-luminosity-metallicity relations ({\plz}) in the near- (e.g \citealt{Longmore1986};  \citealt{Catelan2004a}; \citealt{Sollima2006, Sollima2008}; \citealt{Borissova2009}; \citealt{Muraveva2015,Muraveva2018a}; \citealt{Clementini2017}) and mid-infrared  (e.g. \citealt{Madore2013}; \citealt{Dambis2014}; \citealt{Klein2014}; \citealt{Neeley2015,Neeley2017}; \citealt{Sesar2017a}; \citealt{Clementini2017}; \citealt{Muraveva2018a,Muraveva2018b}) passbands, 
  thus 
 allowing estimations of the distance to the host systems. 

A significant contribution to the study of variable stars and of RRLs in particular, is being provided by the European Space Agency (ESA) mission {\gaia}, which is  designed to chart a three-dimensional map of the Milky Way (MW,  \citealt{Prusti2016,Brown2016}) by repeatedly monitoring the whole sky down to a limiting magnitude of about 21~mag in the {\gaia} $G$-band. {\gaia} Data Release 2 (DR2), on 2018 April 25 
published a catalogue of more than half a million sources classified  as variables of different types in the MW and beyond \citep{Holl2018}. Classification of candidate RRLs in {\it Gaia} DR2 was performed by  (i) the classifiers of the general variability detection pipeline applied to sources with  more than  20 epochs (hereafter {\it nTransits}:20+ classifier; \citealt{Eyer2017}, \citealt{Holl2018}) and, (ii)  by a fully statistical approach specifically developed to classify all-sky high-amplitude pulsating stars  with two or more epoch data (hereafter {\it nTransits}:2+ classifier, \citealt{Rimoldini2019}). The two classification procedures provided a total sample of 228,904 candidate RRLs \citep{Holl2018}. The Specific Objects Study  pipeline for the processing of Cepheids and RRLs (SOS Cep\&RRL; \citealt{Clementini2016,Clementini2019}) confirmed as bona-fide RRLs 140,784 of them, among which approximately 1/3 are new discoveries, 
and provided  their pulsation properties (period, amplitude), along with intensity-averaged mean magnitudes in the {\it Gaia} $G$, $G_{BP}$ and $G_{RP}$ bands calculated by modelling the light curves,  as well as metallicity and extinction for a fraction of them computed from the  Fourier parameters  of the $G$-band light curves \citep{Clementini2019}.  {\gaia} DR2 also published accurate positions, parallaxes and proper motions for a sample of about 1.3 billion sources brighter  than  $G=21$~mag \citep{Brown2018}, which includes a large number of RRLs.  Unfortunately, the accuracy of the {\it Gaia} DR2 parallaxes ($0.02-0.04$~mas for $G<15$~mag) drops dramatically for fainter objects, reaching  values of about 2~mas at $G=21$~mag \citep{Brown2018}, which hampers an accurate estimation of distance directly from {\gaia} parallaxes for sources with such faint magnitudes. Thus,  the use of standard candles such as RRLs becomes crucial to  overcome {\it Gaia}'s limits  in the context of distance scale measurements.

The Draco dwarf spheroidal (dSph) galaxy is a MW satellite located at $\sim76$~kpc \citep{Mc2012} from us. Due to its large distance, the mean {\it Gaia} DR2 parallax of Draco members  happens to be negative  ($\varpi=-0.052\pm0.005$, \citealt{Helmi2018}), hence, basically useless for a direct estimation of distance. However, the HB of Draco is at magnitude  $G\sim20$ mag. This is well above 
{\gaia}'s  limiting magnitude, thus, classification, basic properties and photometry of Draco RRLs are available in the {\it Gaia} DR2 catalogue, and an accurate distance to Draco dSph can be estimated using  the RRL $G$-band luminosity -  metallicity relation ({\gz};  \citealt{Muraveva2018a}). In past years, the RRLs of Draco have been analysed in a number of different studies. \citet{BS1961} discovered 133 RRLs in this dSph. Their photometry was later re-analysed by \citet{Nemec1985} who provided new estimations of period for the RRLs in the \citet{BS1961} sample. \citet{Bonanos2004} provided a catalogue of 146 RRLs observed with the 1.2~m telescope of the Fred Lawrence Whipple Observatory,  of which 131 were already known from  \citet{BS1961}. Finally, \citet{Kinemuchi2008} performed a CCD survey of the Draco dSph galaxy with the 1.0~m telescope at the US Naval Observatory (USNO) and the 2.3~m telescope at the Wyoming Infrared Observatory (WIRO) and presented a catalogue of 270 RRLs, which includes  165 RRLs previously known in this dSph.  

 In this study we have compiled the most complete catalogue of RRLs in Draco  by looking for additional RRLs belonging to this dSph in the variable star catalogues of the Catalina Sky Survey \citep{Larson2003}, the  All Sky Automated Survey (ASAS, \citealt{Pojmanski1997}),  the Lincoln Near-Earth Asteroid Research (LINEAR; \citealt{Stokes2000}), the Palomar Transient Factory  (PTF,  \citealt{Law2009}), Pan-STARRS \citep{Kaiser2010}, the General Catalogue of Variables Stars (GCVS, \citealt{Samus2017}) and in the lists of RRLs published in {\gaia} DR2 \citep{Holl2018, Clementini2019, Rimoldini2019}.
 We have analysed the  Oosterhoff properties and measured individual distances to each RRL in the sample using, for the first time the {\it Gaia} bands, and have studied their  spatial distribution which suggests that Draco may be in tidal interaction with the MW. 

The paper is organised as follows. In Section~\ref{sec:data} we describe our updated catalogue of RRLs in Draco and present the main properties of the Draco RRL population. In Section~\ref{sec:dist} we measure the  distance and analyse the structure of the Draco dSph  as traced by its RRLs. In Section~\ref{sec:oo} we discuss the  Oosterhoff classification of Draco RRLs. In Section~\ref{sec:cand} we present a catalogue of additional candidate RRLs belonging to Draco which were selected based on the uncertainty of their $G$-band  magnitudes in the {\gaia} DR2 catalogue.  A summary of our results and main conclusions are presented in Section~\ref{sec:summ}. 

\vspace{-3mm}

\section{Data}\label{sec:data}

\subsection{Catalogues of RRLs in Draco}\label{subsec:cat}


\begin{figure}
   \includegraphics[trim=10 30 0 0,width=\linewidth]{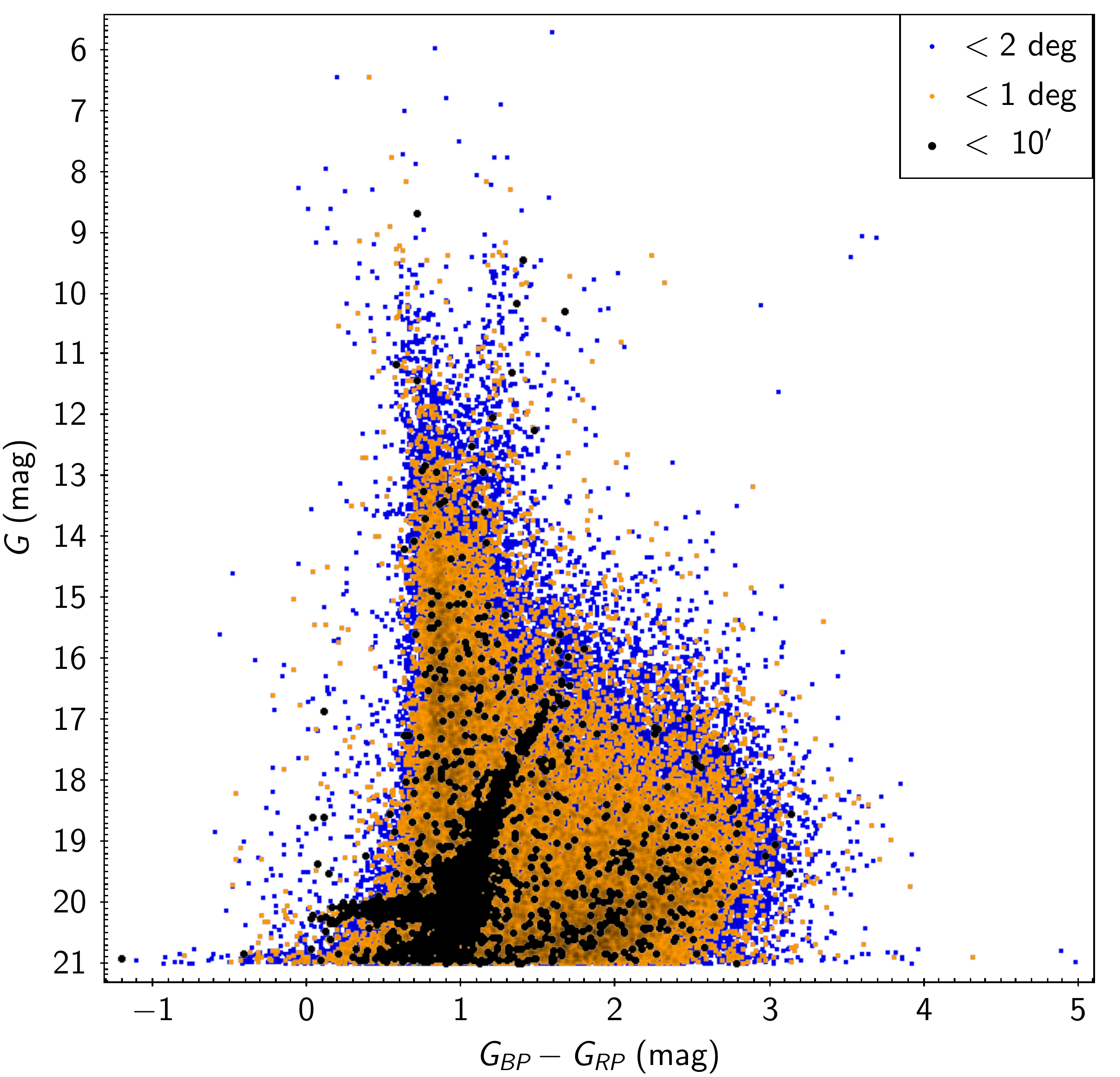}
  \caption{CMD in the {\gaia} passbands of 83,724 sources located within 2~deg (blue circles),  22,221 sources located within 1 deg (orange circles) and 1803 sources located within $10\arcmin$ (black circles)  from the centre of the Draco dSph, according to  \citet{Kinemuchi2008}'s centre coordinates of the galaxy.   
  }
  \label{draco_area_cmd}
\end{figure}

Our main goal was to compile the most complete as possible catalogue of RRLs in the Draco dSph. \citet{Kinemuchi2008} published mean $V$ and $I$ magnitudes, $V$ amplitudes, periods and photometric metallicities  for 9 Anomalous Cepheids (ACs), 2 eclipsing binaries (EBs),  12 slow irregular red variables and 270 RRLs in this dSph, which we have  used as a starting point to build our own catalogue of RRLs in Draco. As a first step,  we searched for RRLs in the field of Draco in the catalogues of currently available large variability surveys (Catalina, ASAS, LINEAR, PTF, Pan-STARRS, GCVS). We selected from these catalogues all RRLs  located  in a circular region of 2 deg in radius around the centre of Draco (RA=260.05162 deg; Dec=57.91536 deg, J2000, \citealt{Kinemuchi2008}).  Such a rather large radius, significantly exceeding the half-light radius of Draco ($10\arcmin$, \citealt{Mc2012}), was adopted in order to include RRLs located in the outskirts of the galaxy and find those which might have been stripped from Draco as a result of the interaction with the MW.

A number of catalogues produced by the Catalina Sky Survey \citep{Larson2003} comprise variable stars located within 2~deg from the centre of Draco. Specifically,  \citet{Drake2014} found 35 periodic variable stars of different types, while 8 RRLs located towards Draco were identified by \citet{Drake2013a}, one by \citet{Drake2013b} and 6 by \citet{Abbas2014}.  No RRLs or  periodic variables of other types were identified in the 
Draco area by the ASAS survey \citep{Pojmanski1998,Pojmanski2000,Pojmanski2002, Pigulski2009} and, similarly, the PTF catalogue does not contain RRLs  belonging to Draco.  On the other hand, 
 \citet{Palaversa2013} and \citet{Sesar2013} found, respectively, 11 periodic variables  and 3 RRLs  in the data of the LINEAR survey \citep{Stokes2000} and  \citet{Sesar2017b} identified 312 RRLs using the multi-band, multi-epoch photometry provided by Pan-STARRS \citep{Kaiser2010}.  Finally, 156 variable stars of different types  are included in the GCVS \citep{Samus2017}.

As a last step, we checked the lists of variable stars published in the {\it Gaia} DR2 catalogue and available through the {\it Gaia}  Archive website\footnote{\url{http://archives.esac.esa.int/gaia}}.
In total 269 DR2 sources in the  Draco area are classified as candidate variable stars by the  {\it nTransit:}20+ and the {\it nTransit:}2+  classifiers of the {\it Gaia}  general variability processing pipeline (\citealt{Eyer2017,Rimoldini2019}). 
Furthermore, the  SOS Cep\&RRL pipeline confirmed the classification as RRLs and provided characteristic parameters for 239 of the variables identified as candidate RRLs  in the Draco
region by the 
classifiers  (\citealt {Clementini2019};  see {\it gaiadr2.vari\_rrlyrae} table).

The {\it Gaia} archive provides three independent measurements of the mean $G$  magnitude of the sources observed by {\it Gaia}: (i) \textsf{phot\_g\_mean\_mag} which is available for all sources in the {\gaia} DR2 general catalogue (see \textit{gaiadr2.gaia\_source} table, hereafter, DR2 \textit{gaia\_source} catalogue) and it is calculated by the  {\it Gaia} photometric processing pipelines \citep{Evans2018}; (ii) \textsf{mean\_mag\_g\_fov} given in table {\it gaiadr2.vari\_time\_series\_statistics} for all stars classified as variables and calculated as the mean magnitude of the time series data \citep{Holl2018}; and (iii) \textsf{int\_average\_g} for RRLs and Cepheids confirmed by the SOS Cep\&RRL pipeline \citep{Clementini2016, Clementini2019}, which is computed as the mean in flux of the Fourier model that best fits the source time-series data ({\it gaiadr2.vari\_rrlyrae} and {\it gaiadr2.vari\_cepheid} tables, for RRLs and Cepheids, respectively), 
%
 with the latter values to be preferred, whenever available \citep{Brown2018,Arenou2018}. 
Since \textsf{phot\_g\_mean\_mag} mean magnitudes are available for all sources in the DR2 general catalogue
and $G_{BP}$ (\textsf{phot\_bp\_mean\_mag}), $G_{RP}$ (\textsf{phot\_rp\_mean\_mag}) mean magnitudes  are available for $\sim82$\% of them 
\citep{Brown2018} we used these mean values in our study of the Draco CMD (Section~\ref{subsec:cmd}). However, we relied on the intensity-averaged magnitudes computed by the SOS Cep\&RRL pipeline, which provide a more accurate estimation of the mean $G$ magnitudes and are available for about 75\% of the RRLs in our sample, to measure the distance and analyse the structure of Draco (Section~\ref{subsec:g}).
 For the remaining stars (18\% of our sample) we either adopted the \textsf{phot\_g\_mean\_mag} mean magnitudes or obtained the $G$-band mean magnitudes by performing our own analysis of the time series data available in the {\it Gaia} archive (6\% of the sample), or transformed the literature  $V$ and $I$ mean magnitudes to $G$ mean magnitudes (1\% of the sample).

Compiling and crossmatching  all the aforementioned catalogues  we obtained a total sample of  379 variables of different types. Among them 336 are classified as RRLs in at least one of the catalogues we have analysed. In order to obtain the most complete census of the RRLs in Draco we proceeded  with the full sample of 336 RRLs, even though for some of them there is  inconsistency of classification among the various catalogues.

\subsection{Sample selection\label{subsec:cmd}}  

In order to extract from the sample of 336 RRLs the true members of the Draco dSph, we applied the following selection procedure:
\begin{enumerate}
\item We constructed the $G$, ($G_{BP}-G_{RP}$) CMD of Draco using sources from the {\gaia} general catalogue.
\item We crossmatched our sample of 336 RRLs against the {\gaia} general catalogue and retrieved their $G$, $G_{BP}$ and $G_{RP}$ magnitudes, which were used to place the sources on the CMD.
\item Based on the distribution on the CMD we selected a sample of RRLs that we suggest  are most likely bona-fide Draco members.
\item To reduce the chances of removing RRLs that are true Draco members but which have incorrect {\gaia} mean magnitudes, we plotted on the CMD  279 RRLs  which have a counterpart in the \citet{Kinemuchi2008} catalogue,  using the  $G$, $G_{BP}$ and $G_{RP}$ magnitudes inferred from their $V$ and $I$ magnitudes, and update our sample based on this. 
\item Finally, we used the {\gaia} proper motions to check the membership to Draco of the RRLs in our sample.
\end{enumerate}

\par\noindent In the following we describe in detail the various steps of our selection procedure. 

We retrieved from the {\gaia} general catalogue all sources (83,724 in total) located within a circular area of 2 deg in radius around the centre of Draco. They are  plotted  as blue points in the $G$, ($G_{BP}-G_{RP}$) CMD in Fig.~\ref{draco_area_cmd},
 whereas orange and black points show sources within 1~deg (22,221 sources) and $10\arcmin$  (1803 sources), respectively. The latter corresponds to the half-light radius of the Draco galaxy according to \citet{Mc2012}. 
The CMD of the 1803 sources within $10\arcmin$ is characterised by a well pronounced Red Giant Branch (RGB) and an HB with mean magnitude approximately at $G\sim20$~mag. We used this CMD in the following analysis to select from our sample of 336 RRLs those which are true members of Draco.

We then crossmatched our sample of 336 RRLs against the DR2 general catalogue and found counterparts within $10\arcsec$ for  335 of them\footnote{One star observed only by the Catalina Sky Survey (J172209.3+560415; \citealt{Drake2014}) has no counterpart in the {\it Gaia} catalogue. \citet{Drake2014} provide only a $V$ mean magnitude for this object  of $V=17.74$~mag, 
 which would place the star $\sim 2$~mag above the HB of Draco ($G\sim20$~mag), hence ruling out that the star can be an RRL belonging to Draco.}
Among these 335 sources  174 (52\% of the sample) are located within $10\arcmin$ from the centre of Draco.
%
%
\begin{figure*}
    \includegraphics[trim=5 20 0 0,width=18cm]{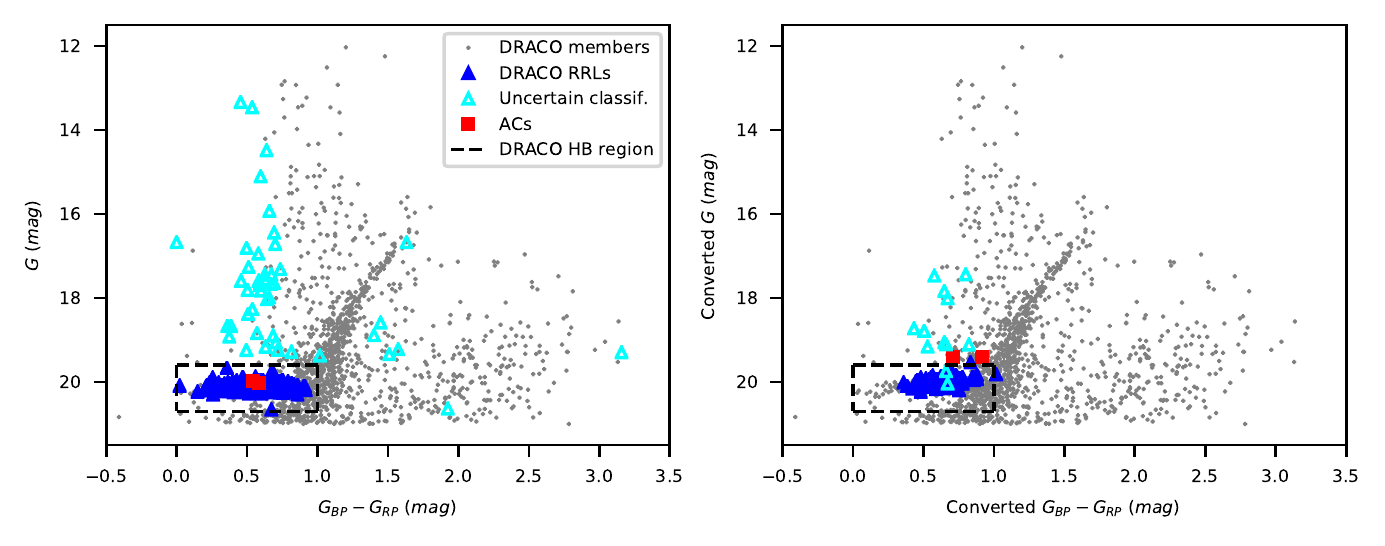}
  \caption{{\it Left panel}: Distribution of the 335 RRLs in our sample on the $G$, $G_{BP}-G_{RP}$ CMD, using $G$, $G_{BP}$ and $G_{RP}$ mean magnitudes from the DR2 \textit{gaia\_source} catalogue; {\it Right panel}: same as in the left panel but for 279 stars  in common with \citet{Kinemuchi2008}, for which $G$, $G_{BP}$ and $G_{RP}$ mean magnitudes were inferred from the \citet{Kinemuchi2008} $V$, $I$ mean magnitudes,  using the transformation relations published by \citet{Evans2018}. See text for details.}
  \label{cmd_rr}
\end{figure*}

The left panel of Fig.~\ref{cmd_rr} shows the distribution of the  335 RRLs  on the CMD of the 1803 sources located within $10\arcmin$ from the  centre. The vast majority of these RRLs are nicely placed on the Draco HB. However, their distribution in colour ($G_{BP}-G_{RP}$) is significantly extended ($\sim1$~mag) with sources showing rather extreme colours, such as  ($G_{BP}-G_{RP}) \sim$ 2.0 - 3.5 mag,  clearly indicating issues with the {\gaia}  $G_{BP}$,  $G_{RP}$ magnitudes of these RRLs. This is not surprising since 
the Draco RRLs  are very close to the {\gaia}  limiting magnitude,  particularly in the  $G_{BP}$ and  $G_{RP}$ passbands. 
We also note that only very few of the RRLs in this region  have  $G_{BP}$, $G_{RP}$ mean magnitudes estimated by the SOS Cep\&RRL pipeline (see {\it gaiadr2.vari\_rrlyrae}), therefore confirming the limited reliability of {\gaia} colours for variable stars with such faint magnitudes.
Additionally,  44 RRLs appear to be significantly brighter than the HB having mean magnitudes between $\sim$ 19.5 and 13 mag in the $G$ band. They might either be foreground RRLs or, more likely, their 
mean magnitudes in the DR2 \textit{gaia\_source} catalogue could be incorrect 
because they are blended with sources not resolved by {\gaia}, or
because the outlier rejection procedure applied for DR2 in the general photometric processing \citep{Evans2018} led to incorrect mean values   (see, \citealt{Brown2018,Arenou2018}).  
Finally, some of these variables could be 
 wrong cross-identifications.
Indeed, crossmatching the different literature lists with the DR2 \textit{gaia\_source} catalogue might have caused wrong cross-identifications due to uncertainties in the source coordinates. 
Whatever the cause, in the following we consider all sources  located in the region: $0<(G_{BP}-G_{RP})<1.0$~mag and $19.6<G<20.7$~mag of the CMD (dashed box in Fig.~\ref{cmd_rr}), as RRLs likely belonging to the Draco dSph and the variables located outside this region (cyan open triangles in Fig.~\ref{cmd_rr}) either as foreground RRLs or as RRLs belonging to Draco  for which the photometry in  the DR2 \textit{gaia\_source} catalogue is incorrect for some of the reasons discussed previously.  
There is a total number of 290 variable stars inside the dashed box in Fig.~\ref{cmd_rr}, among which 288 are classified as RRLs (blue filled triangles) 
and two (red filled squares) are reported as RRLs in all studies, but \citet{Kinemuchi2008} who classify them as ACs. 

\begin{figure*}
    \includegraphics[trim=5 20 0 0,width=18cm]{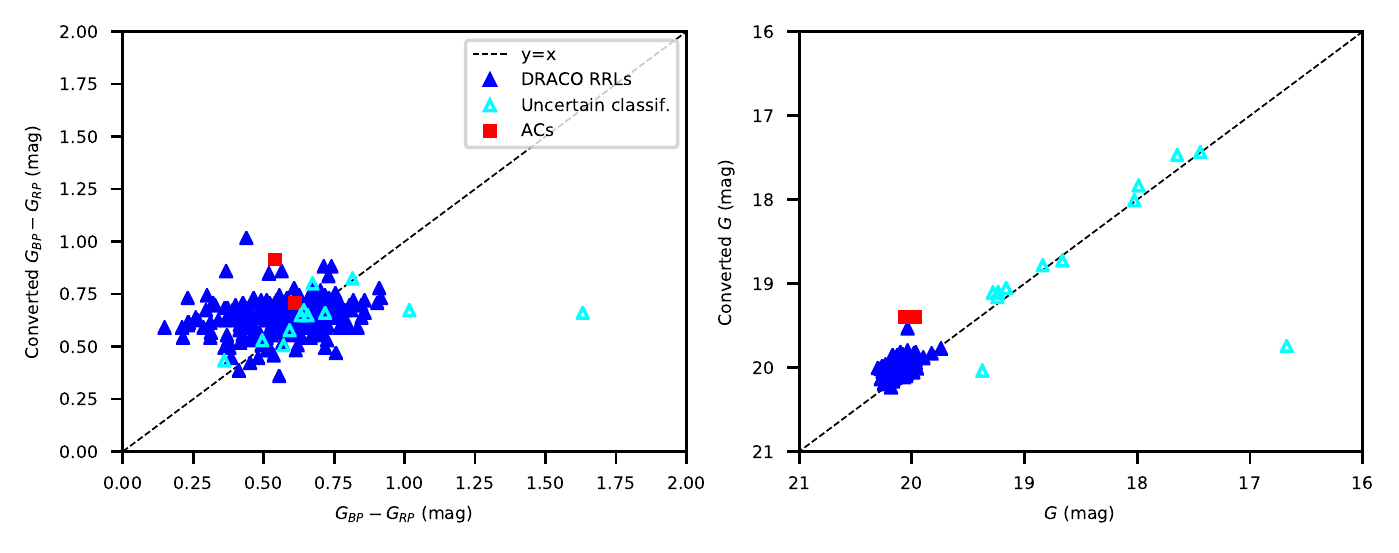}
  \caption{{\it Left panel}: ($G_{BP} - G_{RP}$) colours inferred from the \citet{Kinemuchi2008} $V$, $I$ mean magnitudes using the transformation relations published by \citet{Evans2018} plotted versus ($G_{BP} - G_{RP}$) colours in the {\gaia} general catalogue for  the 279 stars  in common between the two catalogues. The dispersion of the points around the line is 0.17~mag. {\it Right panel}:  same as in the left panel but for the $G$ mean magnitudes. The dispersion of the points around the line is 0.23~mag.}
  \label{conversion_kin}
\end{figure*}

\begin{table*}
\begin{minipage}{18cm}
\caption{Characteristics of our final sample of 285 RRLs in Draco.\label{tab:gen}}
\begin{center}
\tiny
\begin{tabular}{c c c c l c c r r r  c c c}
\hline
\hline

{\gaia} source\_id&RA & Dec & Catalogue$^{*}$ & Type & Period & $Amp(V)$ & $\varpi$~~~~~~~ & $\mu_\alpha\cos\delta$~~~ & $\mu_\delta$~~~~~~& $G$ & Source$^{**}$ & $\mu$ \\
&(deg) & (deg) & & & (days) & (mag)& (mas)~~~~~& (mas/yr)~~~& (mas/yr)~ & (mag) & & (mag) \\
\hline 
 1433901705381651840 & 259.21302 &  58.35194 & (5)           &  RRc  &0.41714   &      0.483 & $  0.13 \pm  0.41 $ &$ -1.00 \pm  0.75 $ & $ 0.27  \pm    0.86 $ &  20.020 & (1) & 19.473  \\     
 1433898406846766848 & 259.24834 &  58.32383 & (5)           &  RRc  &0.48788   &      0.400 & $ -0.30 \pm  0.46 $ &$ -0.30 \pm  0.85 $ & $-0.70  \pm    0.85 $ &  20.052 & (1) & 19.505  \\     
 1433121704960670848 & 259.29090 &  57.86995 & (5)           &  RRab &0.60396   &      0.783 & $ -0.17 \pm  0.54 $ &$ -1.62 \pm  1.21 $ & $ 1.17  \pm    1.30 $ &  20.161 & (1) & 19.615  \\     
 1433128778770683008 & 259.35005 &  58.04222 & (2)           &  RRab &0.55392   &      0.861 & $ -0.36 \pm  0.52 $ &$ -1.40 \pm  1.06 $ & $-0.66  \pm    1.04 $ &  20.130 &  (2)  & 19.584  \\     
 1433055729967360128 & 259.37152 &  57.57381 & (1), (3), (5) &  RRab &0.61233   &      0.700 & $ -0.42 \pm  0.47 $ &$  1.02 \pm  0.88 $ & $-1.60  \pm    1.06 $ &  20.004 &   (3)     & 19.458  \\     
 1433105929544530816 & 259.44539 &  57.72432 & (1), (3), (5) &  RRc  &0.35233   &      0.670 & $ -1.14 \pm  0.47 $ &$  0.31 \pm  0.97 $ & $ 0.15  \pm    1.15 $ &  20.102 &   (3)     & 19.556  \\     
 1433125274077368192 & 259.49618 &  58.03480 & (1), (3), (5) &  RRab &0.62179   &      0.680 & $  0.72 \pm  0.53 $ &$ -0.63 \pm  0.92 $ & $-1.14  \pm    1.00 $ &  20.077 &   (3)     & 19.530  \\     
 1433057314810014464 & 259.50426 &  57.62156 & (2)           &  RRc  &0.37456   &      0.638 & $ -0.40 \pm  0.47 $ &$  -0.12 \pm  0.94 $ & $ 1.57  \pm    1.06 $ &  20.058 &   (2)  & 19.512  \\     
 1433107205150933248 & 259.51113 &  57.80387 & (1), (2), (5) &  RRab &0.62424   &      0.680 & $ -1.07 \pm  0.53 $ &$  -0.92 \pm  1.04 $ & $ 0.46  \pm    1.25 $ &  20.074 &   (2)  & 19.528  \\     

\hline 
\end{tabular}
\end{center}
$^{*}$ The source was classified as RRL by: (1) \citet{Kinemuchi2008}; (2) the {\gaia}  DR2 general variability detection  classifiers  (\citealt{Eyer2017}, \citealt{Rimoldini2019}); (3) the {\gaia} SOS Cep\&RRL  pipeline \citep{Clementini2019}; (4) the GCVS \citep{Samus2017}; (5) Pan-STARRS  \citep{Sesar2017b}.\\
$^{**}$ The $G$ mean magnitude is: (1) taken from the DR2 \textit{gaia\_source} catalogue; (2) calculated in this study with the GRATIS software; (3) taken from table:  {\it gaiadr2.vari\_rrlyrae}, which summarises  results for RRLs obtained by the SOS Cep\&RRL pipeline \citep{Clementini2019}; (4) obtained transforming the $V$, $I$ mean magnitudes of  \citet{Kinemuchi2008} to the {\gaia}  passbands with the transformation relations provided in  \citet{Evans2018}. \\
This table is published in its entirety online (Supporting information); a portion is shown here for guidance regarding its form and content.\\
\normalsize

\end{minipage}
\end{table*}

In order to further test the soundness of our procedure to select bona-fide RRLs belonging to Draco, in the right panel of Fig.~\ref{cmd_rr} we plot the CMD of the sources located  within $10\arcmin$ from the centre of Draco with superimposed  
279 sources, out of our sample of 335 variables, which have a counterpart in the \citet{Kinemuchi2008} catalogue of variable stars in Draco. The $G$, $G_{BP}$ and $G_{RP}$ mean magnitudes of these 279 sources  were computed from the $V$, $I$ magnitudes of \citet{Kinemuchi2008} using the transformation equations provided by \citet{Evans2018}. 
The spread in colour of the RRLs (blue filled triangles) along the Draco HB   is now reduced to less than 0.7~mag. 
Furthermore,  two variable stars classified as ACs by \citet{Kinemuchi2008} (red filled squares) which fell inside the RR Lyrae region in the left panel of Fig.~\ref{cmd_rr} are now located above the HB, consistently with  \citet{Kinemuchi2008}'s classification as ACs. We therefore discard them from our RRL sample. 
Conversely, two RRLs according to \citet{Kinemuchi2008} which were located outside the RRL region in the left panel of Fig.~\ref{cmd_rr}  (cyan open triangles) now nicely fall  within the dashed region of Fig.~\ref{cmd_rr}. They are {\it Gaia} source\_id 1433157331713106304 for which the DR2 \textit{gaia\_source} catalogue provides a $G$ magnitude  about 0.8 mag brighter and a  $G_{BP}-G_{RP}$ colour   about 0.4 mag redder than obtained by transforming to {\it Gaia} passbands the \citet{Kinemuchi2008}'s $V$, $I$ mean magnitudes;  and {\it Gaia} source\_id 1433203652936566016 which  has $V=19.84$~mag in  \citet{Kinemuchi2008} to compare with $G=16.67$~mag, $G_{BP}=17.44$~mag and $G_{RP}=15.81$~mag from the DR2 \textit{gaia\_source} catalogue. For these two RRLs we rely on \citet{Kinemuchi2008} classification and magnitudes, hence we added them  to our sample of bona-fide RRLs in Draco and adopt: $G=20.04$~mag and $G_{BP}-G_{RP}=0.67$~mag for the former  and $G=19.74$~mag and $G_{BP}-G_{RP}=0.66$~mag for the latter, obtained by transforming the \citet{Kinemuchi2008} $V$ and $I$ magnitudes. Our final sample of Draco RRLs thus consists of 290 stars. 

 In the left panel of Fig.~\ref{conversion_kin} we compare the ($G_{BP}-G_{RP}$) colours obtained converting the mean $V$ and $I$ magnitudes of  \citet{Kinemuchi2008}  with  the colours provided  in the {\gaia} main catalogue for the 279 RRLs, while the right panel  shows the same comparison for the $G$ magnitudes.  There is rather poor agreement between the observed and converted colours, again confirming that {\gaia} colours at such faint magnitudes should be treated with caution. Conversely, observed and converted $G$ magnitudes are in good agreement for all but the two  ACs and the two RRLs discussed previously.

\begin{figure}
   \includegraphics[trim=5 30 0 0,width=\linewidth]{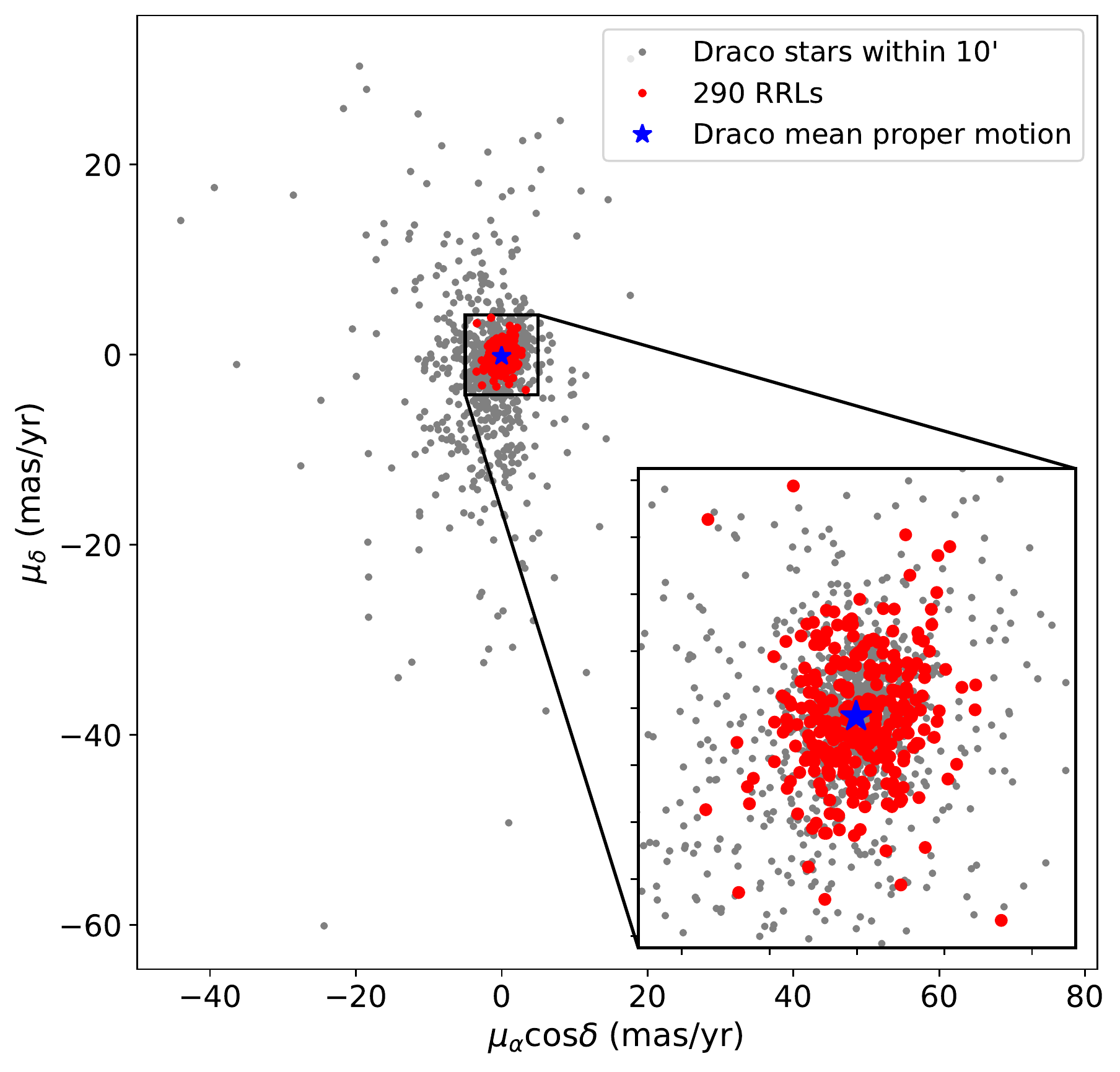}
  \caption{Distribution in the proper motions plane of the 1803 sources located within the half-light radius of the Draco dSph  ($10\arcmin$; grey points). A blue star marks the mean proper motion of the Draco members according to \citet{Helmi2018}. Red circles show the 290 RRLs in our sample.}
  \label{cmd_prmot}
\end{figure}

{\gaia} gives us a further, unprecedented opportunity to check whether these 290 RRLs truly belong to the Draco galaxy through the analysing of their proper motions. In Fig.~\ref{cmd_prmot}  grey points show the distribution in the proper motion plane of  the 1803 sources 
within $10\arcmin$ from the centre of Draco, while red circles mark the 290 RRLs in our sample.  They 
are all within an area of $\pm4$~mas/yr around the mean proper motion value of the Draco members calculated by \citet{Helmi2018}:  $\mu_\alpha \cos \delta = -0.019$~mas/yr; $\mu_\delta = -0.145$~mas/yr (blue star in Fig.~\ref{cmd_prmot}). Their distribution appears to be significantly more concentrated than observed for other sources within the Draco half-light radius (grey points). We consider all 290 RRLs to be true members of Draco based on their distribution on the CMD and proper motion plane. 
An additional test of membership will be performed in Section~\ref{sec:dist} based on the individual distances measured for these RRLs. Among the 290 RRLs  236 (81\%) were classified as RRLs based on the {\it Gaia} DR2 data \citep{Clementini2019,Rimoldini2019}, 267 (92\%) by \citet{Kinemuchi2008}, 131 (45\%) by the GCVS \citep{Samus2017} and  275 (95\%) by \citet{Sesar2017b} using Pan-STARRS data. 

%

\subsection{Mean $G$ magnitudes of  Draco RRLs}\label{subsec:g}

In our study of the Draco CMD (Section~\ref{subsec:cmd}) we relied on the $G$ mean magnitudes estimated in the {\gaia} photometric processing \citep{Evans2018}, since they are available  for all sources in the DR2 \textit{gaia\_source} catalogue.  However, in our study of the distance and structure of Draco using the RRLs (Section~\ref{sec:dist}), a more accurate estimation of the $G$-band mean magnitudes is needed. Following recommendations in \citet{Brown2018} and \citet{Arenou2018} in Section~\ref{sec:dist} we use the $G$-band intensity-averaged magnitudes calculated by model fitting the time-series data as part of the Cepheids and RRLs processing performed with the  SOS Cep\&RRL pipeline \citep{Clementini2019}. These are available for 217 of the 290 RRLs in our sample. 
For the other 19 RRLs which do not have intensity-averaged $G$ magnitudes estimated by the SOS Cep\&RRL  pipeline 
we analysed the time series data available in the {\it Gaia} archive  with the GRaphical Analyzer of TImes Series package (GRATIS, custom software  developed at the Observatory of Bologna by P. Montegriffo, see e.g. \citealt{Clementini2000}) and modelled the $G$-band light curves  adopting the pulsation  periods  from \citet{Kinemuchi2008} and \citet{Sesar2017b} for 16 and one RRLs, respectively, whereas derived the period ourselves  with GRATIS for the remaining two stars.  

For a further 52 RRLs we adopted the $G$ mean magnitudes provided in the DR2 \textit{gaia\_source} catalogue. For the remaining two RRLs 
mean $G$ magnitudes were calculated transforming \citet{Kinemuchi2008} magnitudes, as discussed in Section~\ref{subsec:cmd}.
The mean $G$ magnitudes are provided in column 11 of Table~\ref{tab:gen}. 

\begin{figure*}
 \includegraphics[trim=50 190 50 100,width=\linewidth]{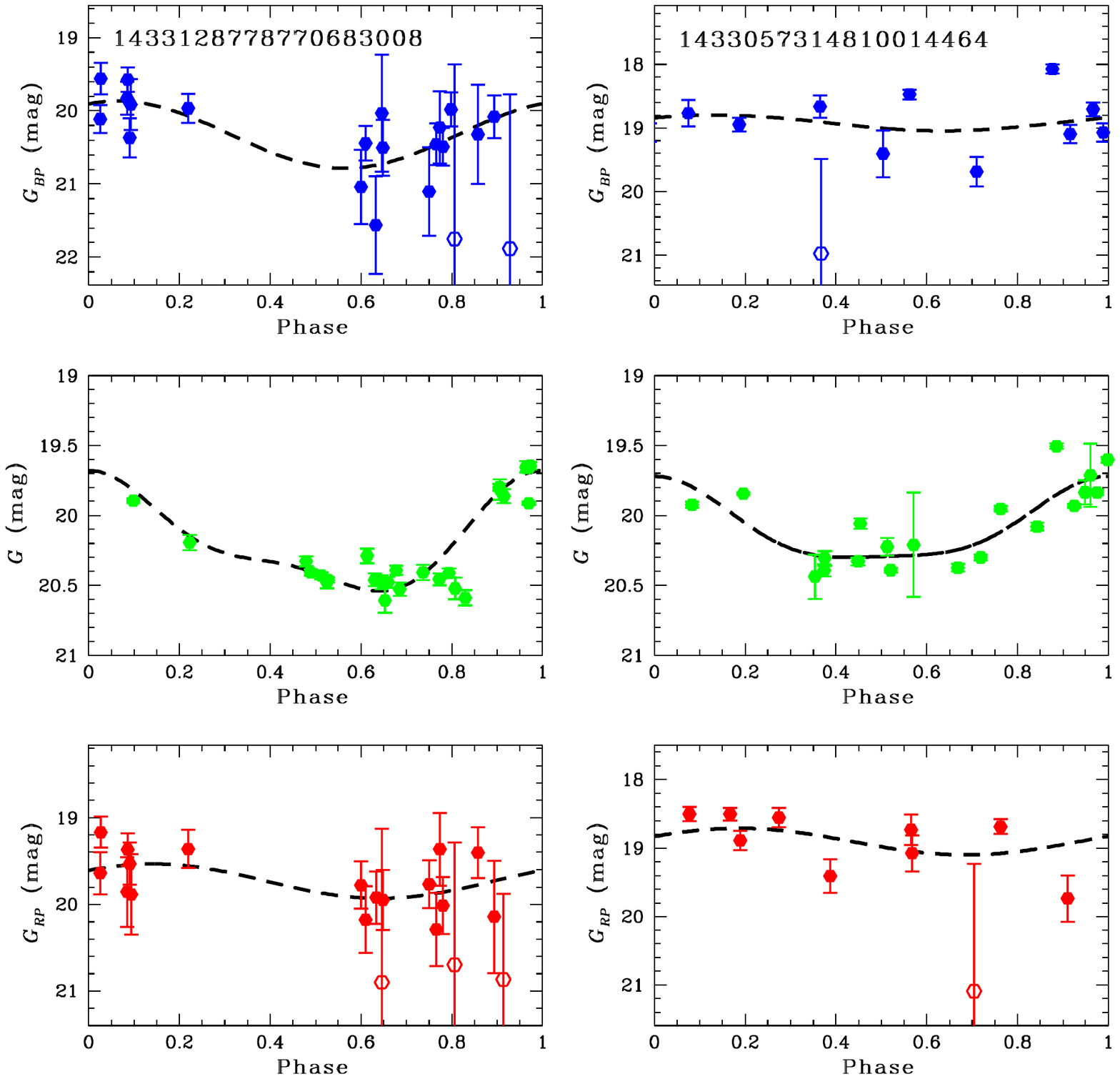}
  \caption{Light curves in the $G$ (green points), $G_{BP}$ (blue points) and $G_{RP}$ (red points) passbands of the two RRLs in Draco discovered by {\gaia}, for which main parameters have been  calculated in the present work. The dashed lines are best-fit models obtained with the GRATIS package. Empty circles mark data points that were discarded during our analysis of the light curves.}
  \label{fig:lc}
\end{figure*}

%
\begin{table*}
\begin{minipage}{15cm}
\caption{New RRLs in Draco discovered by {\it Gaia}.\label{tab:gaia_new}}
\begin{center}
\begin{tabular}{l c c c c c}
\hline
\hline
~~~~~~~~~~~~{\it Gaia}  & Type & Period & $Amp(G)$ &  $G$  & $\sigma_{G}$ \\ 
~~~~~~~~~source\_id    & & (days)  &  (mag) & (mag) & (mag)  \\ 
\hline
1433128778770683008$^{*}$ & RRab  & 0.553924 & 0.861 & 20.130 & 0.134 \\ 
1433057314810014464$^{*}$ & RRc  & 0.374556 & 0.578 & 20.058 & 0.149 \\ 
1433202519064167808$^{**}$ & RRab & 0.551806 & 1.120 & 20.008 & 0.003 \\

\hline 
\end{tabular}
\end{center}
$^{*}$ Parameters derived in this study using the GRATIS package.\\
$^{**}$  Parameters obtained by the {\it Gaia} SOS Cep\&RRL pipeline \citep{Clementini2016, Clementini2019}. \\
\normalsize

\end{minipage}
\end{table*}
The uncertainty in the $G$ mean magnitude of the Draco RRLs which were processed through  the {\it Gaia} DR2 SOS Cep\&RRL pipeline is $\sim$0.005~mag (as estimated via Monte Carlo simulations, see \citealt{Clementini2019})  while is  of $\sim$0.01~mag for the RRLs with $G$ mean magnitude taken from the  DR2 \textit{gaia\_source} catalogue (as calculated from the mean flux uncertainty). 
%
In order to estimate  this uncertainty in a more consistent and rigorous way we have analysed the light curves of a test sample of 75 sources  extracted from the sample of 290 RRLs (25\%) with 
the GRATIS package and estimated the mean dispersion of the data points around the  best-fit models of the light curves  computed with GRATIS:  $\sigma_G=0.1$~mag. We consider this to be a  most reliable estimation of the $G$-band mean magnitude uncertainty and adopt  this value for all the RRLs in our sample. 
%

{\gaia} discovered three new RRLs in Draco: two of them were classified as candidate RRLs by the DR2 general variability detection  classifiers  (\citealt{Eyer2017}, \citealt{Rimoldini2019}) which we  confirm  in our study,  and the third one is a source already confirmed as RRL by the SOS Cep\&RRL pipeline \citep{Clementini2019}. Table~\ref{tab:gaia_new} summarises  information on these three new RRLs. Periods, amplitudes in the $G$-band and intensity-averaged $G$ 
mean magnitudes in the table are those calculated by the {\it Gaia} SOS Cep\&RRL pipeline for  {\it Gaia} source\_id 1433202519064167808, whereas  for the other two sources were derived  in the present study using the GRATIS package. Furthermore, the source with {\it Gaia} source\_id 1433057314810014464 was classified as fundamental mode RRL (RRab)  by the classifiers, however, according to the period derived with the GRATIS package we re-classify this source as first-overtone (RRc) RRL. Fig.~\ref{fig:lc} shows light curves in the $G$, $G_{BP}$ and $G_{RP}$ passbands of the two RRLs, for which main parameters were calculated by us. The quality of the light curves drops dramatically in the $G_{BP}$ and $G_{RP}$ bands, hence no reliable mean magnitudes could be computed in these passbands.

\section{Distance and structure of the Draco dSph}\label{sec:dist}

\begin{figure}
      \includegraphics[trim=5 20 20 0,width=\linewidth]{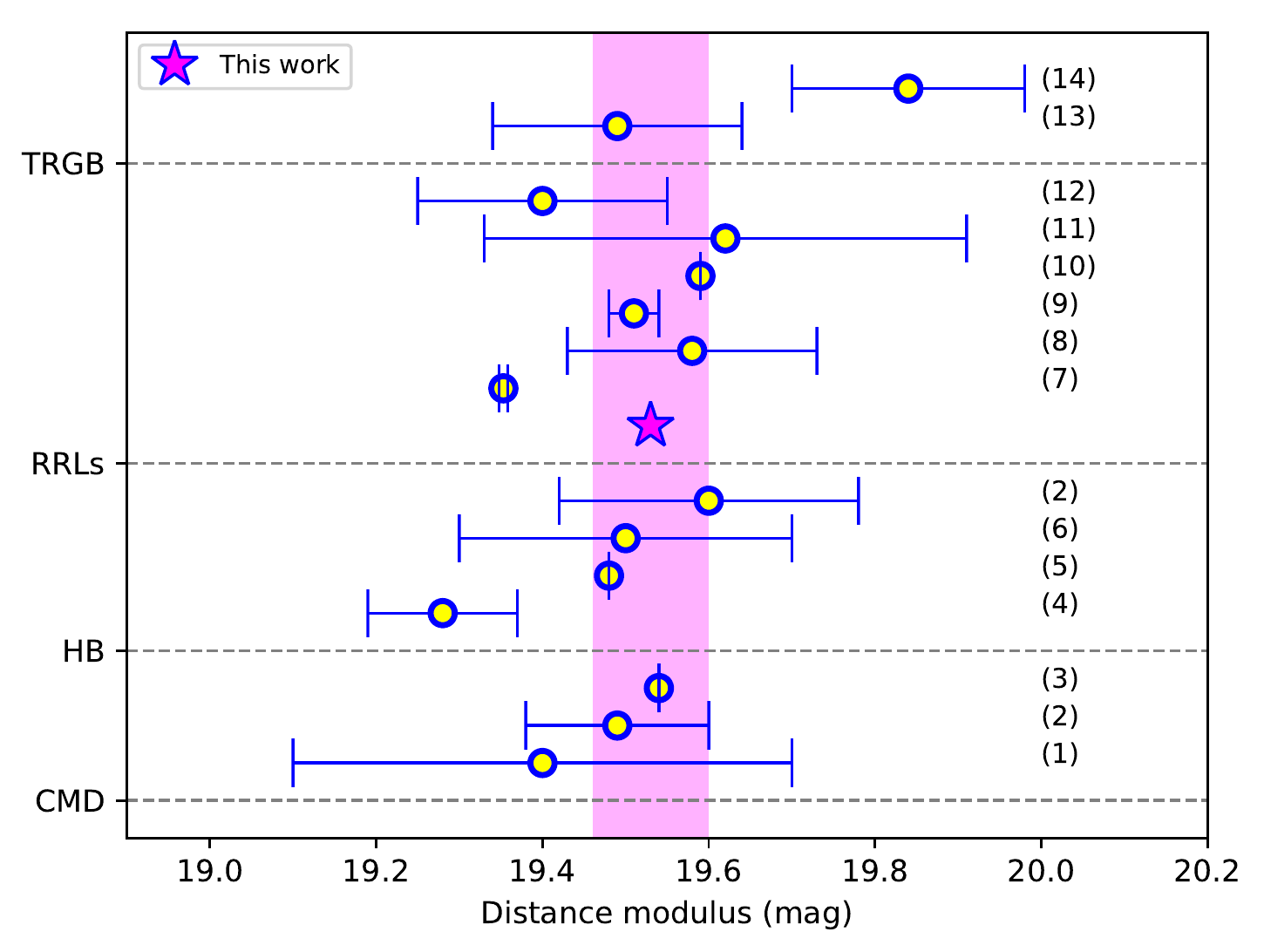}
  \caption{Draco distance moduli estimated using different techniques. The magenta star symbol and shaded region mark the value and uncertainty of Draco distance modulus derived in this work based on a sample of 285 RRLs.  The literature distance moduli of Draco shown in the figure are taken from : (1) \citet{Stetson1979}; (2) \citet{Dolphin2002}; (3) \citet{Weisz2014}; (4) \citet{Irwin1995}; (5) \citet{Grillmair1998}; (6) \citet{Aparicio2001}; (7) \citet{Hernitschek2019}; (8) \citet{Kinemuchi2008}; (9) \citet{Sesar2017b}; (10) \citet{Tammann2008}; (11) \citet{Nemec1985}; (12) \citet{Bonanos2004}; (13) \citet{Cioni2005}; (14) \citet{Bellazzini2002}.}
   \label{fig:dist}
\end{figure}

The most direct method of distance estimation is parallax, however, this technique can be significantly limited for faint distant objects, such as stars in the  Draco dSph. Indeed, \citet{Helmi2018} found a mean value of the {\it Gaia} DR2 parallaxes for Draco members to be negative  ($\varpi=-0.052\pm0.005$~mas), hence, unusable for distance measurement.
Thus, in order to measure the distance to Draco dSph we must rely on indirect techniques.
In the literature there are several estimates of the distance to Draco based on different indirect methods such as:  (i) the galaxy CMD (e.g. \citealt{Stetson1979}; \citealt{Dolphin2002}; \citealt{Weisz2014}); (ii)  the luminosity of the HB (e.g. \citealt{Irwin1995}; \citealt{Grillmair1998}; \citealt{Aparicio2001}; \citealt{Dolphin2002}); (iii)  the RRLs (e.g. \citealt{Nemec1985}; \citealt{Bonanos2004}; \citealt{Kinemuchi2008}; \citealt{Tammann2008}; \citealt{Sesar2017b}; \citealt{Hernitschek2019});  (iv) the tip of the RGB (TRGB; \citealt{Cioni2005}; \citealt{Bellazzini2002}). A comparison of Draco distance moduli obtained by these various studies is presented in Fig.~\ref{fig:dist}.

%

\begin{figure}
   \includegraphics[trim=0 30 0 0,width=\linewidth]{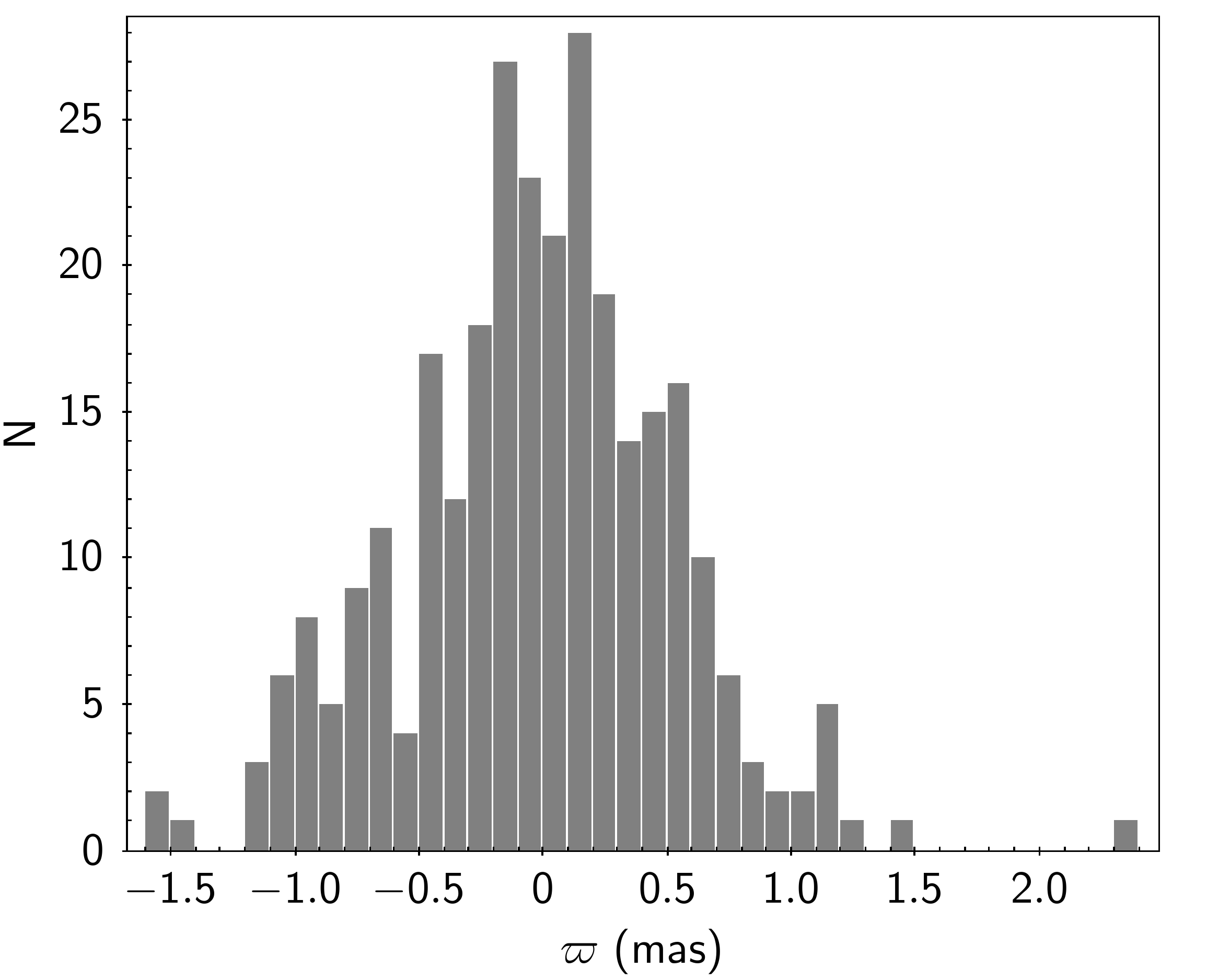}
  \caption{Distribution of the {\gaia} DR2 parallaxes for the 290 RRLs in our sample.}
  \label{fig:hist_par}
\end{figure}

We have used the sample of RRLs selected as described in Section~\ref{subsec:cmd} to  measure the distance and study the structure of Draco.
The mean $G$ apparent magnitude of the 290 RRLs in our sample is $20.08\pm0.08$~mag. At such a faint magnitude, the  uncertainty in {\gaia} DR2 parallaxes  can be as large as 2~mas \citep{Brown2018}, therefore increasing the number of stars with a negative parallax value. This is confirmed by the distribution of parallaxes shown in Fig.~\ref{fig:hist_par}. Only 144 RRLs (50\% of our sample) have a positive value of parallax with a mean relative error $<\sigma_\varpi/\varpi>=3.36$, while the mean parallax of the whole sample of 290 RRLs is $<\varpi_{RRLs}>=-0.02\pm0.48$~mas,  hence cannot be used to measure the distance to Draco.
However, RRLs are valuable tools for indirect measurements of distances because their absolute magnitude can be inferred from a number of fundamental relations these variables conform to 
(Section~\ref{sec:intro}).
In the following, to calculate individual distances to the 290 RRLs in our sample we have used the $M_{G} - {\rm [Fe/H]}$ relation from \citet{Muraveva2018a}:
\begin{equation}
 {\it M_G} = (0.32 \pm 0.04)[Fe/H] + (1.11 \pm 0.06)
\end{equation}
This relation is  calibrated  on {\gaia} DR2  parallaxes of 160 MW RRLs, corrected for the {\gaia}  zero-point offset  \citep{Lindegren2018,Arenou2018} applying a Bayesian approach \citep{Delgado2019},  in combination with accurate $G$-band mean magnitudes computed by the SOS Cep\&RRL pipeline and metallicities from \citet{Dambis2013}. In \citet{Muraveva2018a} we found a non-negligible  dependence of the absolute $G$-band magnitudes on metallicity, hence, an accurate estimation of metal abundance for the Draco RRLs is crucial. \citet{Kinemuchi2008} derived a mean metallicity for  Draco of  ${\rm [Fe/H]} = -2.19\pm0.03$~dex from the Fourier parameters of the light curves of fundamental mode RRLs in this dSph.  \citet{Kirby2013}  measured a metallicity of [Fe/H]=$-1.98\pm0.01$~dex based on spectroscopic observations of 269 Draco members. Finally, \citet{Walker2015} measured individual spectroscopic metallicities for 1565 Draco members, among which 16  RRLs in our sample.
The mean metallicity of these 16 RRLs is ${\rm [Fe/H]}=-1.98\pm0.65$~dex, in excellent agreement with  \citet{Kirby2013} measurement. We therefore have adopted the metallicity estimate by  \citet{Kirby2013} in our analysis, which is based on a larger number of stars. 

\begin{table*}
\caption{RRLs whose membership to  Draco is uncertain \label{tab:cand}}
\begin{center}
\begin{tabular}{l c c c c c c}
\hline
\hline
{\it Gaia}  source\_id &RA  & Dec & Source &  Distance modulus & Angular distance$^{*}$ & Period \\
                                & (deg) & (deg) &         &  (mag)                   & (deg) & (days) \\
\hline

 1432893659376784000 & 261.40951   &      57.29998&  {\gaia} + PS1 & $19.52\pm0.14$ &0.968 &  0.60606 \\
 1422419196214658944 & 263.24290   &      57.79058&  {\gaia} + PS1 & $19.59\pm0.14$ &1.722 &  0.65891 \\
 1433875037928103552 & 258.69214   &      58.01085&  PS1           & $19.33\pm0.14$ &0.708 & 0.47216 \\
 1420713746305299840 & 259.60329   &      56.09359&  PS1           & $20.11\pm0.14$ &1.835 & 0.49933 \\
 1434162354060730752 & 259.04649   &      58.90067&  PS1           & $19.12\pm0.14$ &1.108 & 0.36269 \\
 1432778172000958336 & 258.23426   &      56.31075&  PS1           & $19.35\pm0.14$ &1.873 & 0.49841 \\
 1433085485501417984 & 258.18352   &      57.54400&  PS1           & $19.75\pm0.14$ &1.046 & 0.59077 \\

\hline 
\end{tabular}
\end{center}
$^{*}$  Angular distance from the centre of Draco according to \citet{Kinemuchi2008} \\
\end{table*}

Following  \citet{Bonanos2004} and \citet{Kinemuchi2008} we adopt for the reddening towards Draco the value  $E(B-V)=0.027$~mag \citep{Schlegel1998}, which results in a  $V$-band extinction $A_V=0.084$~mag for a total-to-selective extinction ratio $R_V=3.1$  \citep{Cardelli1989}. The $V$-band extinction was then transformed to the {\gaia} $G$-band extinction $A_{G}=0.070$~mag using relations in  \citet{Bono2019}. By combining the  $G$-band absolute magnitudes \citep{Muraveva2018a} and the  $G$-band apparent magnitudes derived in Section~\ref{subsec:cmd} we obtained individual distance moduli for each of our 290 RRLs.  The uncertainty in these individual distance moduli is on the order of $\sim 0.14$~mag,  due to the combination of the large uncertainties in the mean $G$ magnitudes 
(Section~\ref{subsec:cmd}) and  in the coefficients of the $M_{G} - {\rm [Fe/H]}$ relation from \citet{Muraveva2018a}. Hopefully, both these issues will improve  in {\it Gaia} Data Release 3 (DR3) thanks to a better sampling of the light curves as well as the improved precision and reduced systematics in the parallax measurements. 

\begin{figure}
   \includegraphics[trim=50 40 50 20,width=\linewidth]{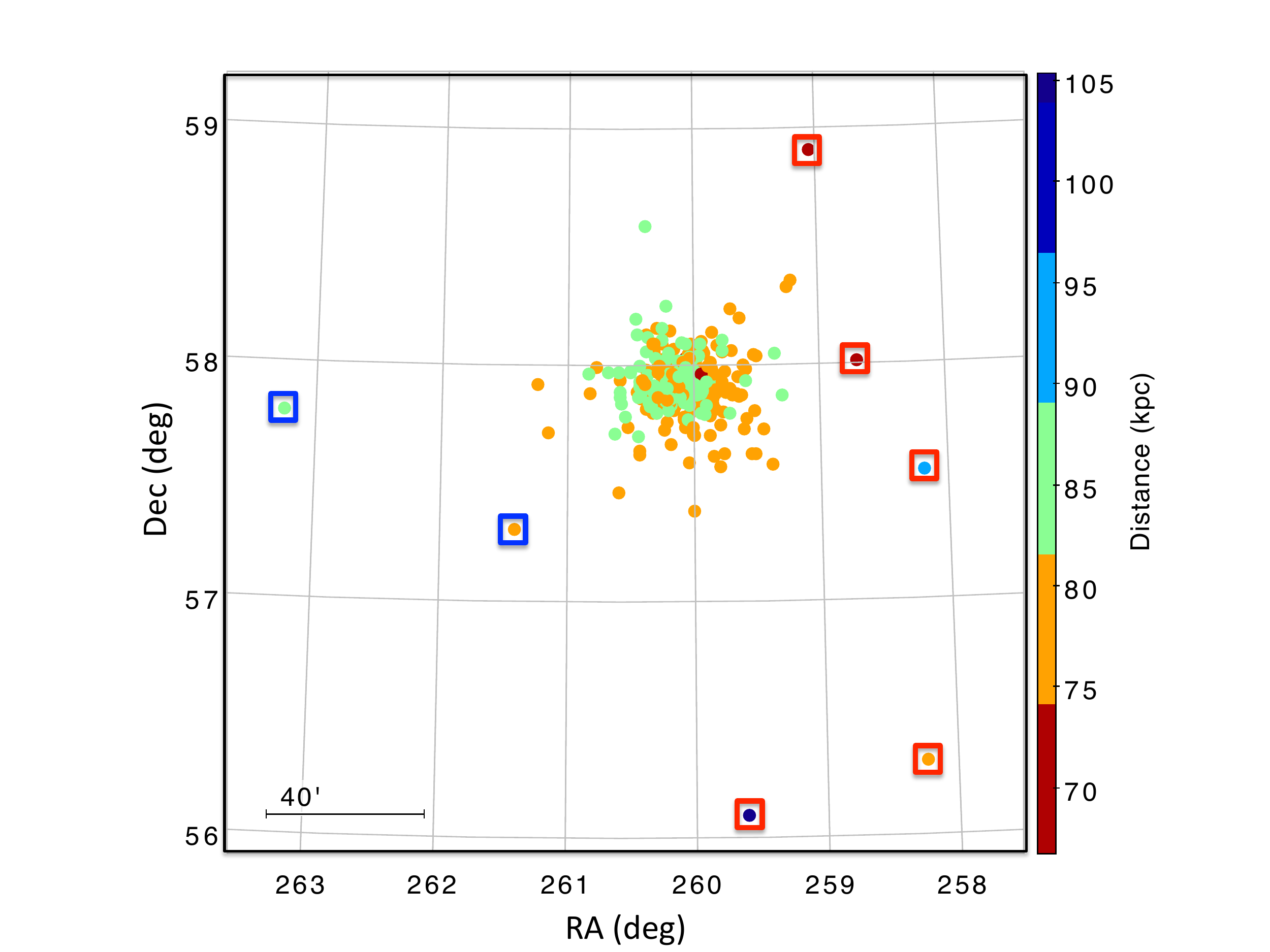}
  \caption{Spatial distribution of the 290 RRLs in our sample. The RRLs are colour-coded according to their distances. Squares highlight sources located at angular distances more than 0.7~deg from the centre of Draco and classified only by Pan-STARRS (red squares) and by Pan-STARRS and {\it Gaia} (blue squares). See text for the details.}
  \label{fig:map_rr}
\end{figure}

Fig.~\ref{fig:map_rr} shows the spatial distribution of the 290 RRLs in our sample, with the RRLs colour-coded according to their distances. Seven sources (highlighted with squares in Fig.~\ref{fig:map_rr}) are located at angular distances more than 0.7~deg from the centre of Draco. All of them are classified as RRLs in the Pan-STARRS catalogue \citep{Sesar2017b}. 
 Coordinates, distance modulus and angular distance from the centre of Draco of these seven RRLs are provided in Table~\ref{tab:cand}. Two of them  (listed in the first two rows of  Table~\ref{tab:cand} and highlighted with blue squares in Fig.~\ref{fig:map_rr})  are RRLs  confirmed by the {\gaia} SOS Cep\&RRL pipeline \citep{Clementini2019}. They have distance moduli of $19.52\pm0.14$ and $19.59\pm0.14$~mag, in good agreement with the mean distance modulus ($\mu = 19.53 \pm 0.07$~mag) derived for Draco using the remaining 283 RRLs, after removing the seven sources under discussion.
 We conclude that they are RRLs belonging to  Draco, perhaps in the process of being stripped away from the galaxy. The remaining five sources (red squares in Fig.~\ref{fig:map_rr}) are classified as RRLs only by Pan-STARRS and their individual distances deviate significantly from the mean distance of the RRL population in Draco. If they are indeed RRLs,  likely they  do not 
belong to Draco, hence we dropped them. 

Our final sample of RRLs belonging to Draco thus consists of  the 285 sources listed in Table~\ref{tab:gen}. The 51 sources discarded from our initial sample of 336 candidate RRLs (Section~\ref{subsec:cat}) 
are listed in Table~\ref{tab:excl}. The distance modulus of Draco based on our final sample of 285 RRLs  is $\mu=19.53\pm0.07$~mag, corresponding to a distance of $80.5\pm2.6$~kpc 
(magenta filled star symbol in  Fig.~\ref{fig:dist}). This value is in good agreement  with estimates of the distance to Draco available in the literature.

\begin{table*}
\begin{minipage}{18cm}
\caption{Characteristics of 51 candidate RRLs discarded from the sample of Draco RRLs.\label{tab:excl}}
\begin{center}
\begin{tabular}{c c c c}
\hline
\hline

{\gaia} source\_id&RA & Dec & Catalogue$^{*}$ \\
&(deg) & (deg) & \\

\hline
  1433066145263630848 &258.04534 &57.22915 &         (2),            (3),                     (5),      (6),    (7)\\ 
  1433153827020071168 &259.77654 &57.83003 &(1),      (2),            (3),           (4),        (5),     (6)      \\ 
  1434263375987484544 &259.67879 &59.22883 &         (2),            (3),                     (5),      (6),    (7)\\ 
  1420862734426225664 &261.72890 &57.00584 &         (2),            (3),                 (5),      (6)          \\  
  1434193587063385856 &257.53717 &58.85367 &(3),           (5),      (6)                                         \\  
  1420748793238290816 &260.53843 &56.07140 &(3),                   (5),      (6)     \\                             
  1433205710224834944 &259.62730 &57.93461 &(1),       (5),      (6)      \\                                          
  1433125411516401920 &259.55362 &58.04889 &(1),        (5),      (6)      \\                                         
  1433228078414897024 &260.07962 &58.27272 &(1),         (5),      (6)    \\                                          
  1432799616772993792 &258.47379 &56.68078 &(4),          (5),     (6)\\                                              
  1433856694124428928 &257.62200 &57.76638 &(5),      (6) \\                                                          
  1434304745112408960 &260.82793 &59.65609 &(5),      (6)      \\                                                     
  1433145649402070784 &260.50300 &57.83918 &(1),       (3),          (4),        (5)     \\                           
  1433205469706734336 &259.78369 &57.97637 &(1),         (2),            (4),        (5),      (6)     \\            
  1433202480409390848 &259.90019 &57.90431 &(1),       (3),        (4),        (5)     \\                            
  1433156167778147584 &260.07411 &57.95209 &(1),         (4),        (5)    \\                                       
  1433735846627961600 &256.79750 &58.06355 &        (2),            (3),          (5),      (6),      (7) \\          
  1433986917532215040 &260.34591 &58.54830 &        (2),           (3),             (5),      (6)\\                  
  1434215272353862272 &257.96432 &59.11785 &         (2),             (3),           (5)\\                          
  1434401257322242816 &261.81705 &57.55481 &        (2),            (3),           (5),      (6)\\                    
  1434768425486064256 &261.76622 &58.87592 &  (2),            (3),         (5),      (6)\\                           
  1437222088762466944 &257.75859 &59.26286 & (2),             (3),              (5)\\                                
  1420767351792676864 &260.20092 &56.22475 &  (2),            (3),             (5),      (6)\\                        
  1420725085018633600 &259.79440 &56.32542 &       (2),            (3),               (5)\\                          
  1422386653247036544 &261.89340 &57.38791 &(2),            (3),                (5),      (6)\\                    
  1433058414322440704 &259.48304 &57.66667 &(1),      (2),            (3),           (4),           (5),      (6)\\  
  1436833999813447296 &256.89134 &58.84992 &       (2),                (5),      (6)\\                               
  1433146718850102912 &260.44898 &57.88857 &(1),          (5)\\                                                       
  1433202519065189248 &259.92731 &57.91376 &(1),          (5)\\                                                       
  1433154789094759040 &260.10482 &57.88138 &(1),      (2),            (3),           (4)\\                           
  1435524172226721920 &261.62292 &59.05737 &        (2),            (3)\\                                             
  1433810548995515648 &257.08989 &58.53205 &(3)\\                                                                      
  1433964446264098048 &258.61022 &58.61517 &(3)\\                                                                      
  1434301206059349376 &260.75617 &59.60149 &(3)\\                                                                      
  1432901214224342272 &260.87156 &57.23603 &(2)\\                                                                      
  1433076556264412416 &258.63455 &57.52561 &(2)\\                                                                      
  1433003095143411328 &258.59581 &57.02518 &(2)\\                                                                      
  1433465852100234752 &256.89900 &57.34440 &(5)\\                                                                      
  1432419804225335168 &257.95292 &56.41195 &(5)\\                                                                      
  1432778172000958336 &258.23426 &56.31075 &(5)\\                                                                      
  1433085485501417984 &258.18352 &57.54400 &(5)\\                                                                      
  1420713746305299840 &259.60329 &56.09359 &(5)\\                                                                      
  1433875037928103552 &258.69214 &58.01085 &(5)\\                                                                      
  1434162354060730752 &259.04649 &58.90067 &(5)\\                                                                      
  1433982686989407104 &259.93819 &58.49680 &(5)\\                                                                      
  1434045084273586432 &259.49137 &58.61261 &(5)\\                                                                      
  1433168120671188096 &261.04323 &57.97529 &(5)\\                                                                      
  1422294642162968576 &262.06508 &56.75950 &(5) \\ 
  1420573764731259136 & 261.16735 &  56.01170 & (5)\\                                                                  
  1432934096493482752 &260.50193 &57.48558 &(5),      (6), (7)\\                                                      
   $-$ $^{**}$                   &260.53908 &56.07089 &(6) \\    
  \hline 
\end{tabular}
\end{center}
$^{*}$ The source was included in the catalogue of variable stars of: (1) \citet{Kinemuchi2008}; (2) the {\gaia}  DR2 general variability detection  classifiers  (\citealt{Eyer2017}, \citealt{Rimoldini2019}); (3) the {\gaia} SOS Cep\&RRL  pipeline \citep{Clementini2019}; (4) the GCVS \citep{Samus2017}; (5) Pan-STARRS  \citep{Sesar2017b}; (6) the Catalina Sky Survey; (7) the LINEAR survey.\\  
$^{**}$  No counterpart was found within 10 arcsec in the {\it gaia\_source} catalogue.\\

\normalsize

\end{minipage}
\end{table*}                                                       


 \begin{figure}
   \includegraphics[trim=50 50 20 20,width=\linewidth]{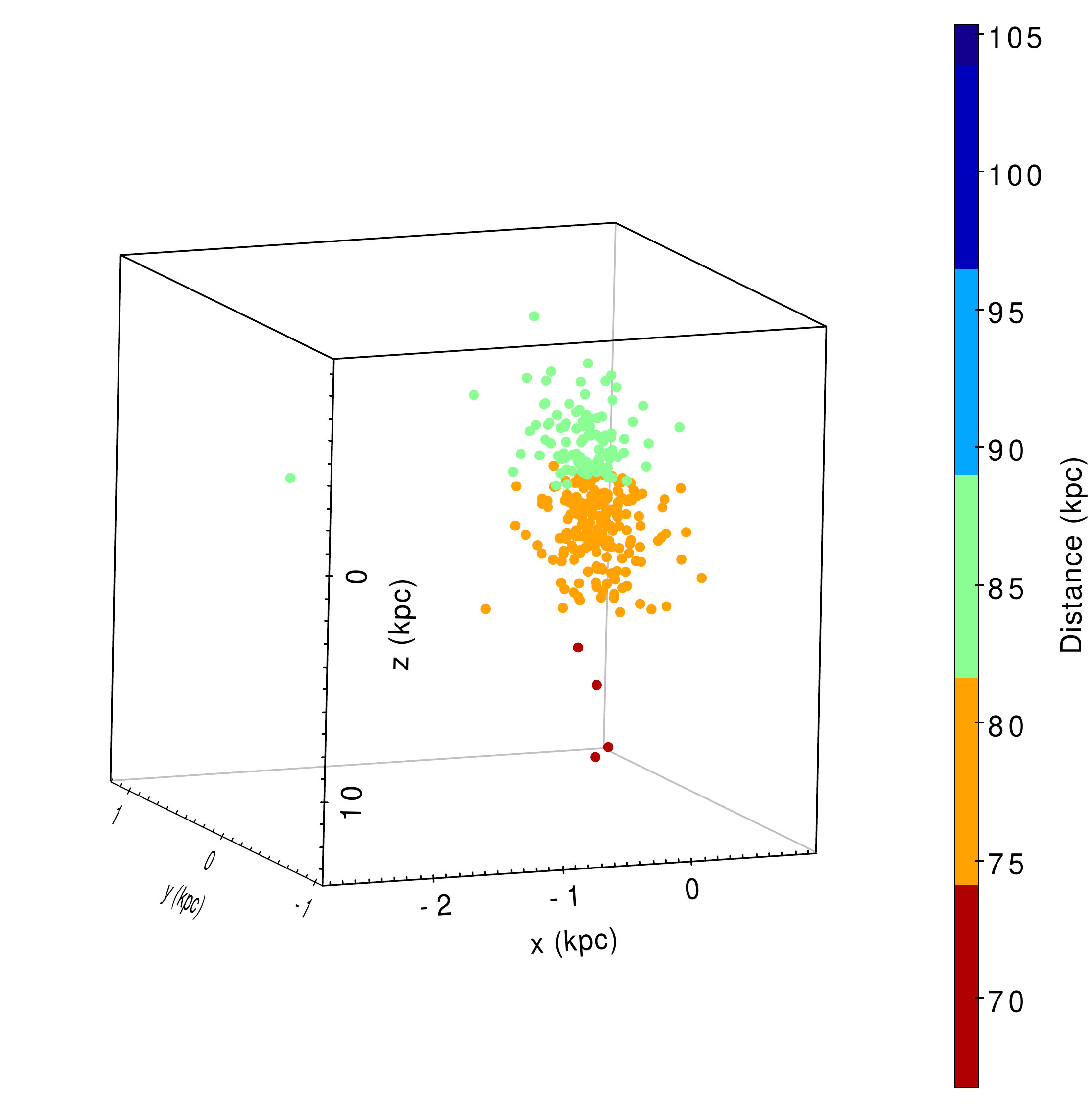}
  \caption{Three-dimensional distribution, in Cartesian coordinates, of our final sample of 285 RRLs in Draco. The colour scale encodes the source distances.}
  \label{fig:3D}
\end{figure}

 \begin{figure}
  \includegraphics[trim=10 40 0 10,width=\linewidth]{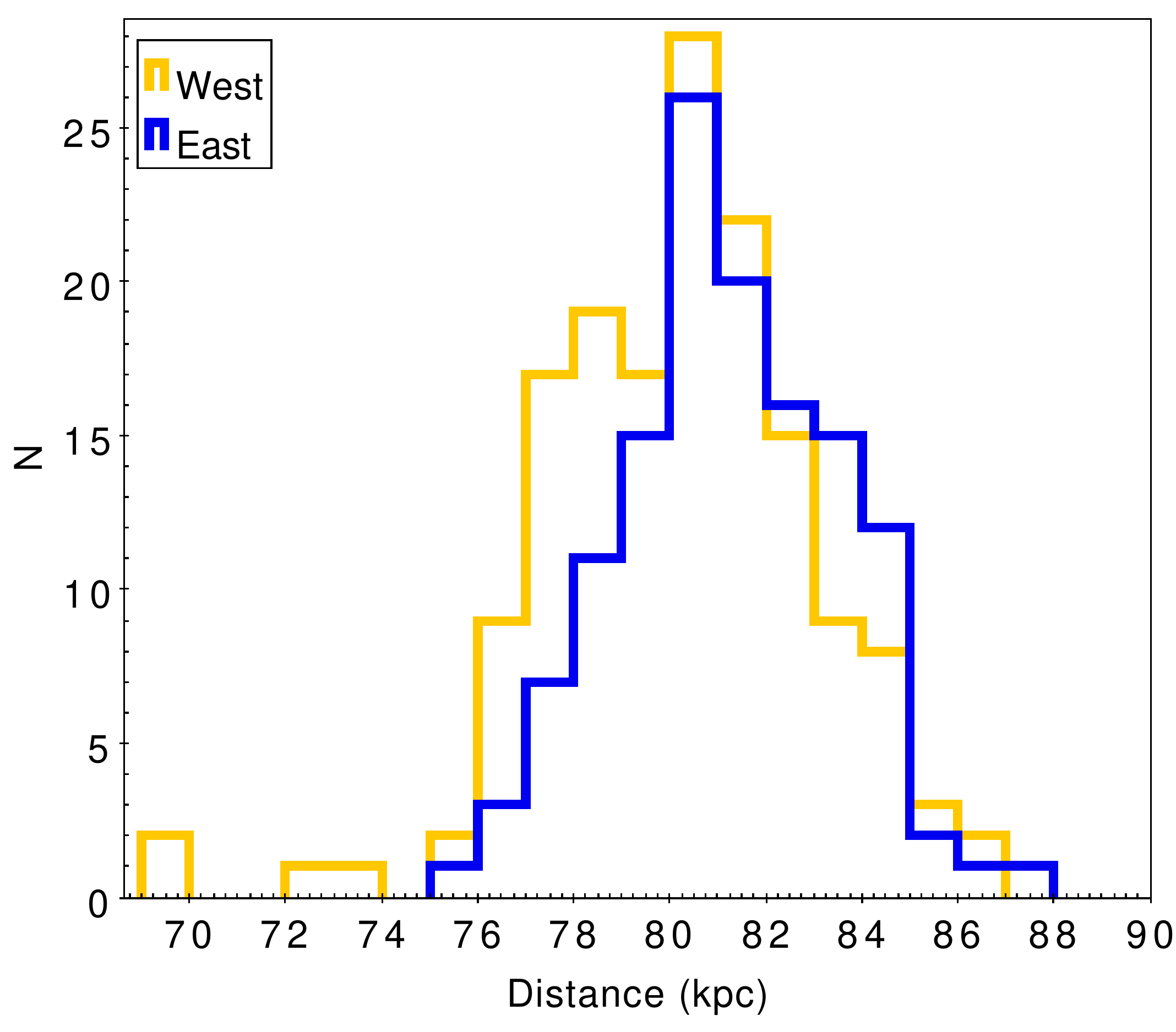}
  \caption{Distance distributions of the RRLs in the eastern (blue line, 155 sources) and western (yellow line, 130 sources) regions of Draco.}
  \label{fig:hist_dist}
\end{figure}

 \begin{figure}
  \includegraphics[trim=70 40 70 30,width=\linewidth]{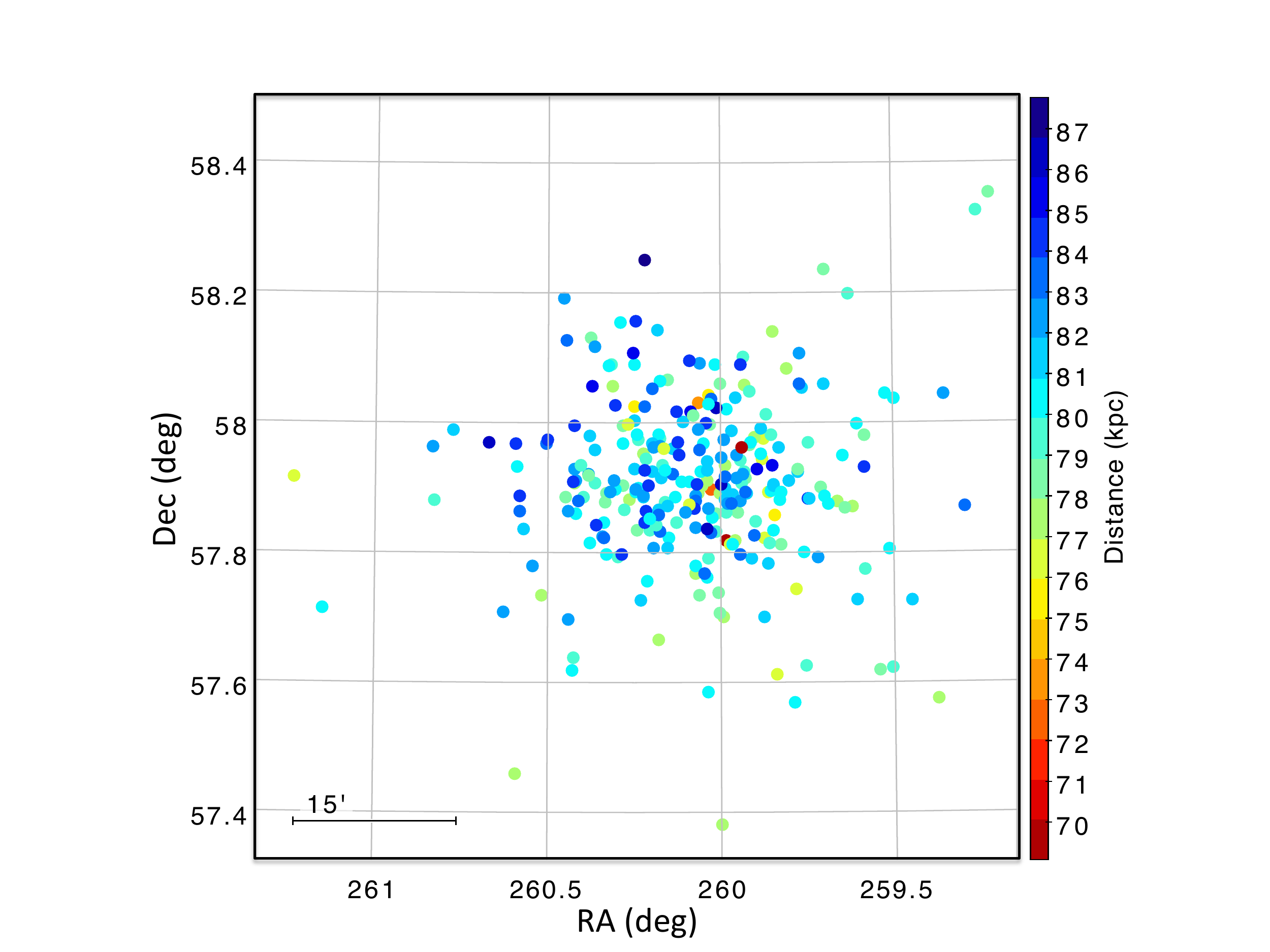}
  \caption{Spatial distribution of the RRLs in the central region of Draco (within $38\arcmin$ from the centre of the galaxy). The RRLs are colour-coded according to their distances.}
  \label{fig:map_rr_zoom}
\end{figure}

The distribution of RRLs in Fig.~\ref{fig:map_rr} seems to suggest that the western/south-western part of the Draco dSph might be closer to us, as if the halo of Draco traced by the RRLs were  tilted likely due to the interaction with the MW. 
 In order to better investigate this possibility, in Fig.~\ref{fig:3D} we show the three-dimensional distribution in $x, y, z$ Cartesian coordinates of the 285 RRLs in our final sample. The Cartesian coordinates were obtained from the RRLs'  $RA$, $Dec$ coordinates and individual distances, using transformation equations from \citet{vanderMarel2001} and assuming  as coordinates  and distance to the origin of the system the centre coordinates of \citet{Kinemuchi2008}: $RA_0 = 260.05162$~deg; $Dec_0=57.91536$~deg, and  the mean distance of $D_0=80.5$~kpc as derived from our sample of 285 RRLs. The $x$-axis was assumed to be antiparallel to the $RA$ axis, the $y$-axis is parallel to the $Dec$ axis and the $z$-axis extends along the line of sight with values increasing towards the observer. 
The three-dimensional distribution of the 285 RRLs also seems to suggest a possible tilt of Draco's halo.

As a further test, we have divided the 285 RRLs into a western sample ($RA<RA_{\rm av}$) and an eastern sample ($RA>RA_{\rm av}$) containing 155 and 130 RRLs, respectively, where $RA_{\rm av}=260.094$~deg is the average right ascension of the full sample of 285 sources. 
The mean distance of the  RRLs is  $81.2\pm2.3$ and $80.2\pm2.8$~kpc, in the eastern and western regions of Draco, respectively, where uncertainties were calculated as the standard deviation of the mean.  These mean values also seem to indicate that the western region of Draco is $\sim1$~kpc closer to us than the eastern region. The distance distributions (adopting a bin size of 1 kpc) of the RRLs in the eastern (blue line) and western (yellow line) regions of Draco are shown in Fig~\ref{fig:hist_dist}. They also seems to indicate that the RRLs in the western region may be located closer to us.  However, errors are still so large that the two distances/distributions cannot be considered statistically different.
Finally, Fig.~\ref{fig:map_rr_zoom} shows a zoom-in of the spatial distribution of the RRLs in the central region of Draco. 
To conclude, the existence of a possible tilt in the Draco's halo remains more a qualitative result, which will require confirmation based on better accuracy data to achieve a statistical significance.

It is worth noticing that the typical uncertainty of the individual RRL distances  is on the order of $\sim$  5~kpc, a main contributor being the uncertainty in the apparent $G$ magnitudes (Section~\ref{subsec:g}). The improvement in photometric accuracy and the increased  number of  epoch-data for variable sources expected with {\gaia} DR3 
will likely allow us to make 
a more sound analysis of the possible tilt of Draco's halo.

\section{The Oosterhoff dichotomy in the Draco dSph}\label{sec:oo}

The Oosterhoff dichotomy \citep{Oo1939} is the observational evidence that the Galactic globular clusters (GCs) can be divided in two separate groups based on the  properties of their RRL population.  The mean period of RRab and RRc stars in Oosterhoff type I (Oo~I) clusters is  $<P_{ab}>=0.55$ and  $<P_{c}>=0.32$ days, respectively, and the fraction of RRc stars over total number of RRLs is $\sim 17$\%. Clusters of Oosterhoff type II (Oo~II) instead contain RRLs  with  $<P_{ab}>=0.64$ and  $<P_{c}>=0.37$ days and the fraction of RRc stars is  $\sim44$\%. Oo~I GCs are also more metal-rich than the Oo~II GCs. Lately, field MW RRLs were also found to exhibit the Oosterhoff dichotomy, while systems outside the MW do not necessary show it. In particular, the vast majority of the classical dSphs around the MW have Oosterhoff intermediate (Oo-Int) properties (e.g. \citealt{Catelan2004b}, \citealt{Clementini2010}), implying that systems like the classical dSphs have not  provided a major contribution to the stellar content of the MW halo through hierarchical merging. 
The Draco dSph is known to belong to the Oo-Int class (e.g. \citealt{BS1961}; \citealt{Bonanos2004}; \citealt{Kinemuchi2008}), even though  \citet{Kinemuchi2008} also found 
that RRc and double-mode RRLs (RRd) in Draco show the characteristic properties of the Oo~II systems. It is clear though that in order to fully investigate the Oosterhoff type of a system one needs a sample of its RRL population as complete as possible. We have thus re-analysed the Oosterhoff class of Draco using our enlarged sample of 285 RRLs. 

For 267 RRLs in our sample the period and $V$-band amplitude ($Amp(V)$) are available from \citet{Kinemuchi2008}. For other 10 RRLs,  $G$-band amplitudes ($Amp(G)$), periods and classification in type were provided by the SOS Cep\&RRL pipeline \citep{Clementini2019}.  Three others sources were classified as RRLs by the {\it nTransit:}2+ classifier  \citep{Rimoldini2019} and their $G$-band time series photometry is available on the {\gaia} archive. For two of them (first two entries in Table~\ref{tab:gaia_new}) we determined the period, classification in RRL type and $G$-band amplitude using the GRATIS software, while for the third source we adopted the period and classification in type from Pan-STARRS \citep{Sesar2017b} and estimated the $G$-band amplitude with GRATIS. To transform the $G$-band amplitudes to amplitudes in the $V$ band we then used equation~2 in \citet{Clementini2019}.  Finally, for five RRLs observed only by Pan-STARRS we took periods and amplitudes in the Sloan $g$ band  ($Amp(g)$) from \citet{Sesar2017b} and  transformed the RRL amplitudes from the Sloan $g$ to the Johnson $V$ band, following \citet{Marconi2006}. 
%
%
In their figures~11 and 12 these authors show that the ratio between $g$ and $V$ amplitudes for RRab and RRc stars is independent of period and metallicity and approximately equal to $Amp(g)/Amp (V)\sim1.2$. 
In their catalogue \citet{Sesar2017b} only provide a probabilistic score for an RRL to be an RRab or an RRc pulsator.  For the 5 RRLs observed only by Pan-STARRS we adopted a classification based on these scores. The characteristics (classification in type, period and $V$-band amplitude obtained as described above) for  our sample of 285 RRLs in Draco are summarised in Table~\ref{tab:gen}. 
Our final sample is composed by 224 RRab, 35 RRc and 26 RRd stars.

\begin{table*}
\caption{Candidate RRLs in Draco selected from the {\gaia} DR2 catalogue based on the dispersion of their $G$-band  magnitudes ($\sigma_{G}$).\label{tab:candidate}}
\begin{center}
\begin{tabular}{c c c c c}
\hline
\hline

{\gaia} source\_id & RA & Dec & $G$ & $\sigma_G$\\
  & (deg) & (deg) & (mag) & (mag) \\
\hline 
  1432983162200486528 & 261.59588  &57.96386 & 20.225 &  0.010\\
  1432870462257494400 & 259.94842  &56.95765 & 20.003 &  0.009 \\
  1432864414943731712 & 260.79201  &57.10239 & 19.977 &  0.013\\
  1432954884135224832 & 260.97300  &57.50792 & 20.185 &  0.011\\
  1432949077339507712 & 260.65013  &57.61801 & 20.096 &  0.017\\
  1432843597237448064 & 259.78716  &57.25623 & 20.215 & 0.011\\
  1432885717981606784 & 261.08232  &57.19960 & 19.923 & 0.013\\
\hline 
\end{tabular}
\end{center}

This table is published in its entirety online (Supporting information); a portion is shown here for guidance regarding its form and content.\\
\normalsize

\end{table*}

\begin{figure}
   \includegraphics[trim=10 40 0 10,width=\linewidth]{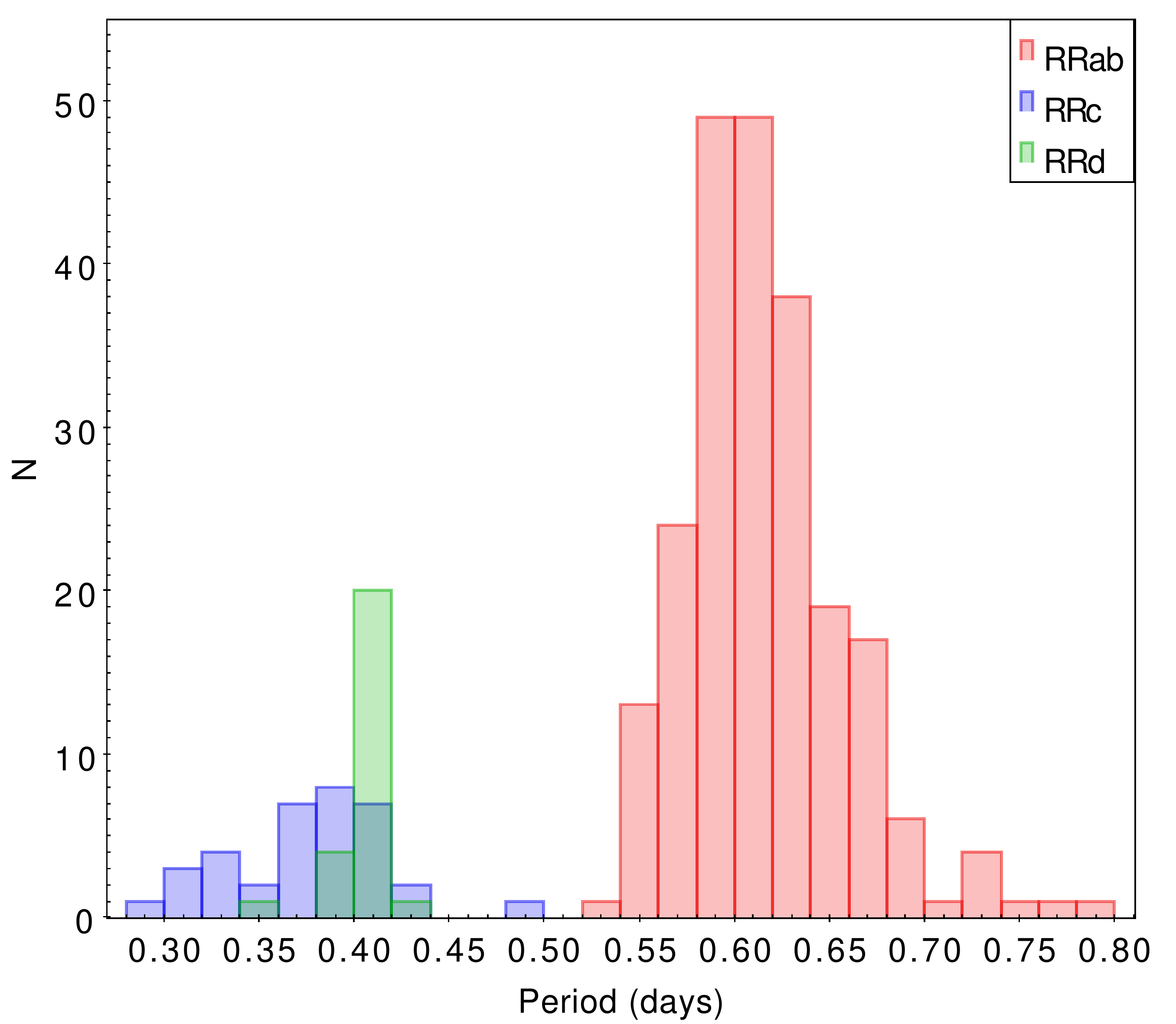}
  \caption{Period distribution of the Draco RRab (red), RRc (blue) and RRd (green) stars in our final sample of 285 sources.}
  \label{fig:hist_per}
\end{figure}

\begin{figure}
   \includegraphics[trim=20 170 20 90,width=\linewidth]{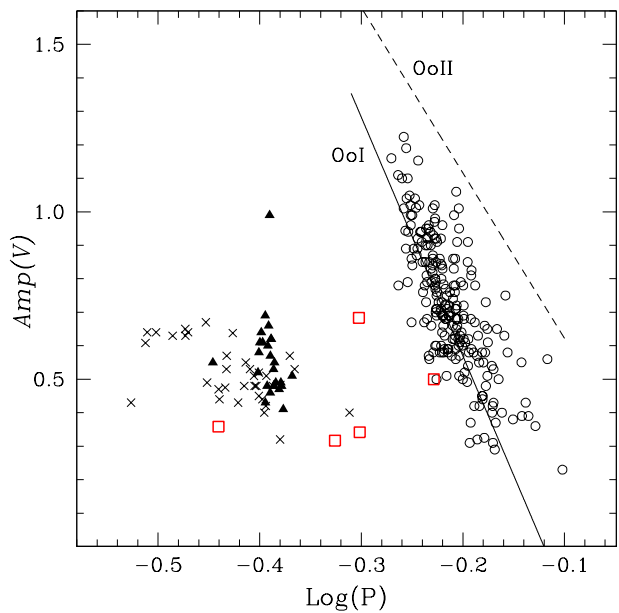}
  \caption{Period-amplitude diagram of the Draco RRLs.  RRab, RRc and RRd stars (285 in total) are shown with empty circles, crosses and filled triangles, respectively. Five RRLs that we discarded based on their distance moduli and angular distances from the centre of Draco (Section~\ref{sec:dist}) are marked by red empty squares. See text for details. The Oo~I and II lines are from \citet{Clement2000}.}
    \label{fig:ooster}
\end{figure}

Red, blue and green histograms in Fig.~\ref{fig:hist_per} show the period distributions of our sample of RRab, RRc  and RRd stars in Draco. The first-overtone period is shown for the RRd stars. As expected these distributions are very similar to the one in figure~5 of \citet{Kinemuchi2008}. In Fig.~\ref{fig:ooster} the period-amplitude diagram of the RRLs in Draco is compared with the Oo~I and Oo~II  loci of Galactic GCs by \citet{Clement2000}. In the figure we have marked with red empty squares the 5 RRLs that we  discarded  based on the distance moduli and angular distance from  the centre of Draco (see Section~\ref{sec:dist}, last five entries in Table~\ref{tab:cand} and red squares in Fig.~\ref{fig:map_rr}).  These five RRLs deviate from the bulk of RRL distribution on the period-amplitude diagram, thus endorsing our decision to discard them from the sample. 

The mean periods of  RRab and RRc stars 
are $<P_{ab}>=0.615\pm0.042$ and $<P_c>=0.377\pm0.040$ days, respectively, and the ratio of number of RRc and RRd stars over total number of RRLs is 21\%.  The distribution of the RRab stars suggests an Oo~I/Oo-Int classification for Draco.  An Oo-Int class is also confirmed by the $<P_{ab}>$ value. However, the mean period of the RRc stars is  more typical of an Oo~II system and the percentage of  RRc and RRd stars is more similar to an  Oo~I. To summarise, based on our enlarged sample of 285 RRLs we re-confirm the Oo-Int nature of Draco,  as already reported in the literature (e.g. \citealt{BS1961}, \citealt{Bonanos2004}, \citealt{Kinemuchi2008}).

\section{Candidate RRLs in Draco}\label{sec:cand}

\begin{figure}
  \includegraphics[trim=10 30 0 10,width=\linewidth]{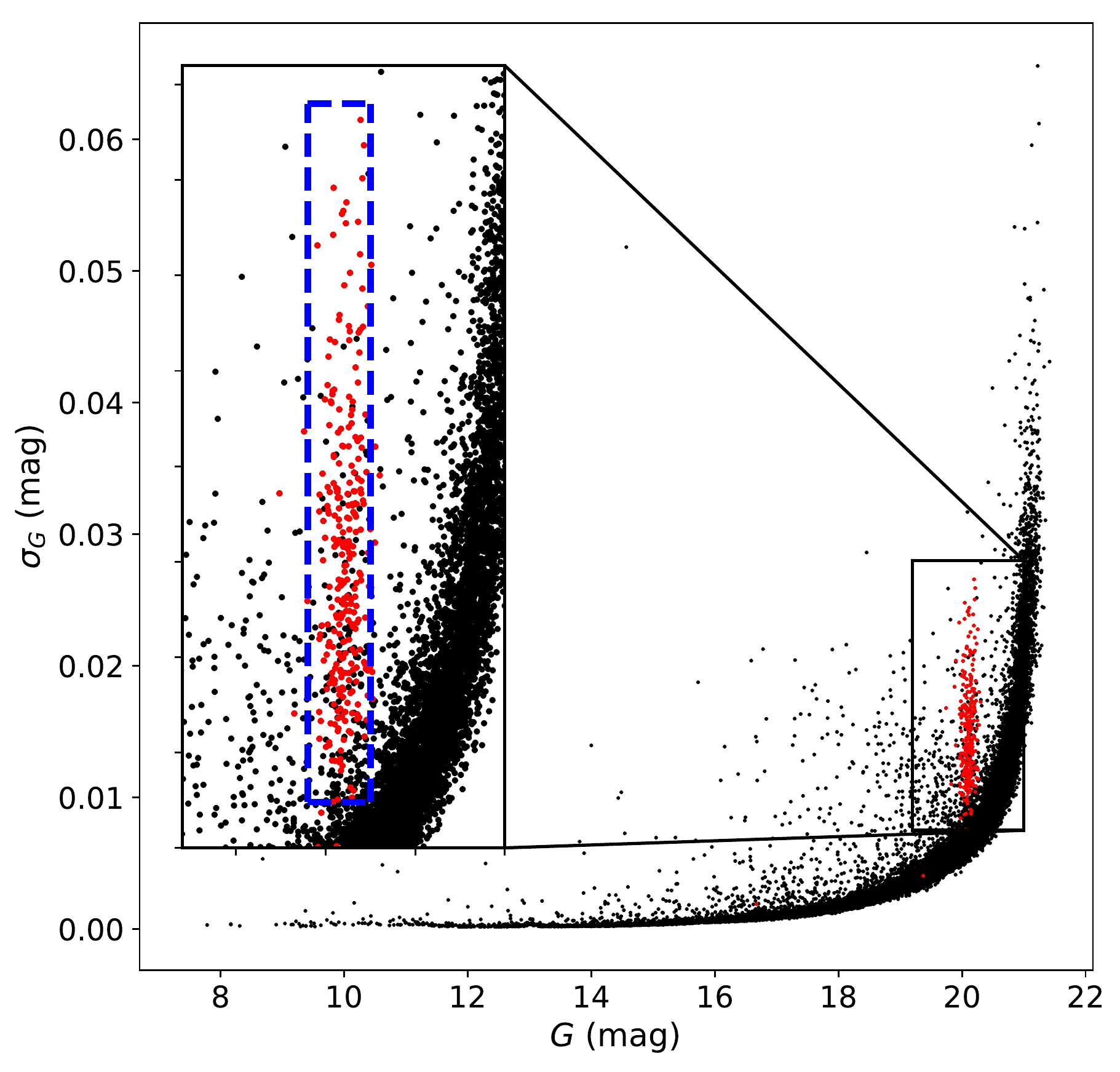}
  \caption{Distribution of sources in the {\gaia} general catalogue located within 1~deg from the centre of Draco (black dots) in the $\sigma_G$ versus $G$ plane (scatter diagram). The 285 confirmed Draco RRLs in our sample are shown with red dots. Blue dashed lines outline the region populated by candidate Draco RRLs. See text for the details.}
  \label{fig:cand_distr}
\end{figure}


The variability of a source causes its mean magnitude, as estimated from a  sequence of observations, to carry a larger dispersion than for a constant star of the same magnitude. We have used such an effect to search for additional RRLs in the  {\gaia} general catalogue of sources located within 1 deg from the centre of Draco.
 Fig.~\ref{fig:cand_distr} shows the distribution of these sources (black points) 
  in the $\sigma_G$ versus $G$ plane (scatter diagram), where the $G$-band magnitudes are taken from the {\gaia} general catalogue and the $\sigma_{G}$ values are calculated from the uncertainties in flux. In the figure 
the characteristic vertical feature (``finger'') at $G\sim 20$~mag  corresponds to the RRLs in Draco. All stars located in a  box with $19.9<G<20.25$~mag and $0.0087<\sigma_{G}<0.027$~mag  (blue dashed lines in Fig.~\ref {fig:cand_distr}) may be potentially  RRLs belonging to  Draco. This sample consists of 448 stars, which we further selected as to have a colour in the range $0<G_{BP}-G_{RP}<1$~mag, corresponding to the colour distribution of the RRLs in Draco  (see Fig.~\ref{cmd_rr}).  For 22 among the 448 sources in the above box an information on the colour is missing in the {\it Gaia} general catalogue, therefore, we exclude them from our analysis. Of the remaining 426 sources, 312 meet the selection in colour and 
269 of them are already included in our sample of 285 RRLs in Draco (Section~\ref{sec:dist}). They were marked as red circles in Fig.~\ref {fig:cand_distr}. The remaining 43 stars are potentially new RRLs of Draco.  Two of them were classified as ACs by \citet{Kinemuchi2008} (red filled squares in Fig.~\ref{cmd_rr}) that we dropped from the list of RRLs  based on the analysis of the Draco CMD (Section~\ref{subsec:cmd}). We consider the remaining  41 sources (Table~\ref{tab:candidate}) as  candidate RRLs belonging to Draco. 
More epoch data which will become available in {\gaia} DR3,  may help shedding light on the actual nature of these stars. 


\section{Summary}\label{sec:summ}

Aiming to collect a sample of RRLs in the Draco dSph as complete as possible we performed an extensive  search for RRLs in the literature and in the databases  produced by large surveys (Catalina, ASAS, LINEAR, PTF, Pan-STARRS, GCVS), as well as in  the catalogue of variable stars published in {\gaia} DR2 \citep{Brown2018}. Combining different catalogues we have obtained a sample of 336 sources located within 2~deg from the centre of Draco, which have been classified as RRLs in at least one of the datasets we have analysed. From this sample we retrieved a subset of  285 RRLs which we consider to be true members of Draco based on: (i) an analysis of their location on the galaxy $G$, ($G_{BP}-G_{RP}$) CMD; (ii) a study of their proper motions; (iii) an investigation of their distances and spatial distribution. Three among these 285 RRLs are new discoveries by {\gaia}.

We determined individual distances to these 285 RRLs applying the {\gz} relation from \citet{Muraveva2018a} and used them to measure the distance and study the structure of the Draco dSph. The mean distance modulus of Draco from the RRLs is: $\mu=19.53\pm0.07$~mag, corresponding to a distance of $80.5\pm2.6$~kpc, in very good agreement with previous estimates available in the literature. 
There is some indication that the  RRLs populating the western/south-western part of Draco may be located closer to us, hence, the halo of Draco might be tilted as a result of interaction with the MW. However, the  large uncertainty in the individual RRL distances ($\sim 5.2$~kpc) does not allow us to obtain a statistically robust proof of such an effect. A new full investigation will be carried out  when more epoch data and more accurate parallaxes will become available with {\gaia} DR3. 

We re-evaluated the Oosterhoff classification  of Draco using the period-amplitude diagram and the mean period of the RRab stars defined by our enlarged sample of RRLs and confirm the intermediate Oosterhoff nature of the Draco dSph already reported in previous studies.
Finally, we used the dispersion in the mean magnitude of sources in the {\gaia} general catalogue with $G \sim$ 20 mag   located  within 1~deg from the Draco centre 
to identify a sample of further 41 candidate RRLs in this dSph.

This study shows once again the  great potential of {\gaia} in the field of variable stars and, at the same time,  how variable stars such as the RRLs allow us to  extend our capability to measure distances well beyond the reach of {\gaia} astrometry.  A further significant contribution to both topics will be achieved with {\gaia} DR3 currently foreseen for the second half of 2021.




\section*{Acknowledgements}
We warmly thank the referee Dr V. Scowcroft  for her valuable comments and suggestions which have significantly improved our paper. 
This work makes use of data from the ESA mission {\it Gaia} (\url{https://www.cosmos.esa.int/gaia}), processed by
the {\it Gaia} Data Processing and Analysis Consortium (DPAC, \url{https://www.cosmos.esa.int/web/gaia/dpac/consortium}). 
Funding for the DPAC has been provided by national institutions, in particular
the institutions participating in the {\it Gaia} Multilateral Agreement.
Support to this study has been provided by the Agenzia Spaziale Italiana (ASI) through grants ASI 2014-
025-R.1.2015 and ASI 2018-24-HH.0,  and by Premiale 2015, ``MITiC" (P.I. B. Garilli).

\section*{Data availability}

The data underlying this article are available in the article and in its online supplementary material.





\bsp	
\label{lastpage}
\end{document}